\documentclass[12pt,preprint]{aastex}
 
\usepackage{emulateapj5} 
 
\slugcomment{Accepted by The Astrophysical Journal}

\shorttitle{Molecular Differentiation in Starless Cores}
\shortauthors{Tafalla et al.}

\begin{document}
 
\title{Systematic Molecular Differentiation in Starless Cores}

\author{M. Tafalla}
\affil{Observatorio Astron\'omico Nacional, Alfonso XII, 3, E-28014 Madrid,
Spain}
\email{tafalla@oan.es}
 
\author{P.C. Myers}
\affil{Harvard-Smithsonian Center for Astrophysics, 60 Garden St., Cambridge,
MA 02138, USA}
\email{pmyers@cfa.harvard.edu}
 
\author{P. Caselli and C.M. Walmsley}
\affil{Osservatorio Astrofisico di Arcetri, Largo E. Fermi 5, I-50125
Firenze, Italy}
\email{caselli@arcetri.astro.it, walmsley@arcetri.astro.it}
 
\and
\author{C. Comito}
\affil{Max-Planck-Institut f\"ur Radioastronomie,
Auf dem H\"ugel 69, 53121 Bonn, Germany}
\email{ccomito@mpifr-bonn.mpg.de}

\begin{abstract}
We present evidence that low-mass starless cores, the simplest units of
star formation, are systematically differentiated in their chemical
composition. Molecules including CO and CS almost vanish near the core centers,
where the abundance decreases by at least one or two orders of magnitude
with respect to the value in the outer core. At the same time, the
N$_2$H$^+$ molecule has a constant abundance, and the fraction of
NH$_3$ increases toward the core center. Our conclusions are based on
a systematic study of 5 mostly-round starless cores (L1498, L1495, L1400K,
L1517B, and L1544), which we have mapped in
C$^{18}$O(1--0), CS(2--1), N$_2$H$^+$(1--0), NH$_3$(1,1) and (2,2)
and the 1.2 mm continuum (complemented with
C$^{17}$O(1--0) and  C$^{34}$S(2--1) data for some systems).
For each core we have built a
spherically symmetric model in which the density is derived from
the 1.2 mm continuum, the kinetic temperature from NH$_3$, and
the abundance of each molecule is derived using a Monte Carlo
radiative transfer code which simultaneously fits the shape of the
central spectrum and the radial profile of integrated intensity.
Regarding the cores for which we have C$^{17}$O(1--0) and  
C$^{34}$S(2--1) data,
the model fits these observations automatically when the standard
isotopomer ratio is assumed.
 
As a result of this modeling, we also find that the gas kinetic
temperature in each core
is constant at approximately 10 K. In agreement with previous
work, we find that if the dust temperature is also constant,
then the density profiles are centrally flattened, and we can model them
with a single analytic expression. We also find that for each core
the turbulent linewidth seems constant in the inner 0.1 pc.
 
The very strong abundance drop of CO and CS toward the center of each core
is naturally explained by the depletion of these molecules onto dust grains
at densities of 2-6 $\times 10^4$ cm$^{-3}$. N$_2$H$^+$ seems
unaffected by this process up to densities of several $10^5$ or
even $10^6$ cm$^{-3}$, while the NH$_3$ abundance may be enhanced by
its lack of depletion and reactions
triggered by the disappearance of CO from the gas phase.

With the help of the Monte Carlo modeling, we show that chemical
differentiation automatically explains the discrepancy between the sizes of
CS and NH$_3$ maps, a problem which has remained unexplained for
more than a decade. Our models, in addition, show that a combination
of radiative transfer effects can give rise
to the previously observed discrepancy in the linewidth of these two
tracers. Although this discrepancy has been traditionally
interpreted as resulting from
a systematic increase of the turbulent linewidth with radius, our
models show that it can arise in conditions of constant gas turbulence.
\end{abstract}

\keywords{ISM: abundances--- ISM: clouds---ISM: molecules---stars: formation}
 
\section{Introduction}

Dense molecular cores are the basic units of star formation in
nearby clouds like Taurus and Perseus, where stars like our Sun have
been forming over the last few million years \citep[e.g.,][]{mye99}.
The study
of the physical structure and kinematics of these cores 
is therefore crucial for our understanding of the star formation 
process, and molecular lines play a role in every step of this work.
They probe density and
temperature through their excitation, and turbulent and systematic motions
through their linewidth and Doppler shifts. For this reason, chemical
anomalies in the core gas can hinder our attempt to understand core
properties, as the 
lack of a full theory of core chemical composition has made it a standard
practice to assume a homogeneous abundance for all molecular species.

The presence of chemical inhomogeneities in the
star-forming material at scales of dark clouds has been known 
for some time, with TMC-1 and L134N being 
the most studied  examples \citep[e.g.,][]{lit79,pra97,swa89,dic00}.
The large-scale abundance gradients in these clouds seem
best explained with time-dependent, gas-phase chemistry, implying
that different condensations have evolved with different
time scales (see \citealt{lan00} and \citealt{van98}
for recent reviews). At the smaller size
of the dense cores, a series of recent observations has shown that 
in some cases, the abundance of molecules like CO and CS 
decreases toward the core center 
(L1498: \citealt{kui96,wil98}, IC5146: \citealt{kra99,ber01},
L977: \citealt{alv99}, L1544: \citealt{cas99}, L1689B: \citealt{jes01}).
These decreases in abundance have been interpreted as resulting from 
the depletion of molecules onto dust grains at the high densities 
and cold temperatures occurring in dense core interiors (e.g., 
\citealt{ber97,cha97}).
It is not clear, however, whether these drops in abundance are typical 
of all dense cores, or they are
limited to a small number of objects. To answer this question, we
have carried out a systematic study of a sample of 5 starless cores
by observing them in a similar manner and analyzing their emission with the same
radiative transfer modeling. 

\section{Observations}

Our core sample is listed in Table 1 and was selected by inspecting 
the N$_2$H$^+$(J=1--0) maps of \citet{cas01} and choosing cores with bright
emission and nearly round contours. Our goal was to select cores that had
the simplest internal structure and lacked obvious outside perturbation. 
In order to achieve high spatial resolution, the cores were chosen
from the nearby Taurus complex (estimated distance of 140 pc,
\citealt{eli78}), with the exception of L1400K, which is at an estimated
distance of 170 pc \citep{sne81}. Later mapping showed that L1400K 
deviates significantly from spherical symmetry, which together with its 
non-Taurus membership made a strong case for eliminating it from
the sample. We decided however to retain this source in our sample 
to avoid any possible bias.
As we will show below, this core presents the same chemical behavior as
the others, but its elongated shape makes it the hardest and most
uncertain to model.

The starless classification of the cores in our sample is based on the 
analysis by \cite{ben89}, who found that they are not associated with IRAS or 
NIR sources. Unlike \cite{ben89}, however, we consider L1495 to be 
starless, given the large separation between this core and IRAS 04112+2803 
(aka CW Tau), and the lack of apparent connection between the two. 
According to \cite{mye87}, the lack of IRAS detection in a Taurus core
implies a luminosity limit of $\le 0.1$ L$_\odot$ for a possible
embedded source.

We observed our sample of cores in NH$_3$
with the 100m telescope of the MPIfR in Effelsberg in 1998 October
and 2001 May. The simultaneous (J,K)=(1,1) and (2,2) 
observations were done in frequency switching (FSW) mode with a throw
of 4 MHz, and used the AK90 autocorrelator to provide a velocity 
resolution of 0.03 or 0.06 km s$^{-1}$, depending on configuration.
Cross scans on continuum point sources were used 
to estimate and correct pointing errors, and line
observations of L1551 and S140 were used to
calibrate the data, using as a reference the intensities reported
by \citet{men85} and \citet{ung86}. The telescope
beam size at the observing frequencies is approximately 40$''$ (FWHM).

We observed the same 5 cores in C$^{18}$O(J=1--0), CS(J=2--1), and 
N$_2$H$^+$(J=1--0) with the FCRAO\footnote[1]
{FCRAO is supported in part by the National Science Foundation
under grant AST 94-20159, and is operated with permission of the Metropolitan
District Commission, Commonwealth of Massachusetts.}     
14m telescope in 1999 November, 2000 April, and 2001 April. To
obtain optically thin tracers for testing our radiative transfer models,
L1498, L1517B, and L1544 were also observed in C$^{17}$O(J=1--0), and L1544 
was observed C$^{34}$S(J=2--1). All observations
were done with the SEQUOIA array receiver in FSW mode, using frequency 
throws of 8 MHz for  N$_2$H$^+$(1--0), and 4 MHz for the rest of the lines. 
This resulted in a velocity resolution of  0.06 km s$^{-1}$ for 
N$_2$H$^+$(1--0), and 0.03 km s$^{-1}$ for the rest of the lines.
The pointing was checked and corrected 
using observations of SiO masers, and the data were 
converted into main beam temperature scale using an efficiency of 0.55 
\citep{lad96}. The telescope beam size at the frequencies of
observation is approximately $50''$.

Finally, we observed L1498, L1495, L1400K, and L1517B
in the 1.2mm continuum with the IRAM 30m telescope in 1999 December. The
observations were done in on-the-fly mode with the MPIfR 37-channel
bolometer array \citep{kre98}, using a scanning speed of 
$4''$ s$^{-1}$, a wobbler 
period of 0.5 s, and wobbler throws of 41$''$ and 53$''$. Three maps of L1495
were done and later combined, and the other sources were observed making single
maps. The sky optical depth was estimated from sky dips before and after 
each individual map. All data were reduced using the NIC software 
\citep{bro96}, and a global calibration factor of 15000 counts per
Jy beam$^{-1}$ was estimated from an observation of Uranus. The bolometer
central frequency is 240 GHz and the 
telescope beam size is approximately $11''$.

For the analysis of the very narrow lines presented in this paper it is
necessary to use an accurate set of rest frequencies.
Currently available line catalogs (e.g., \citealt{pic98}) do not
always have the necessary precision, so a combination of accurate
laboratory measurements and astronomical observations is still necessary.
Here we use the set of frequencies recommended by \citet{lee01},
who combine new (unpublished) laboratory determinations
by C. Gottlieb with astronomical observations of the narrow-line
core L1512, which is an improvement of the frequency set used by 
\citet{taf98}. Following \citet{lee01}, we use these
frequencies: 93176.258 MHz for N$_2$H$^+$(JF$_1$F=101--012), 
97980.953 MHz for CS(J=2--1), 109782.172 
for C$^{18}$O(J=1--0), and 23694.4949 MHz for NH$_3$(J,K=1,1).

\section{Results}


\begin{figure*}
\epsscale{1.5}
\figurenum{1}
\plotone{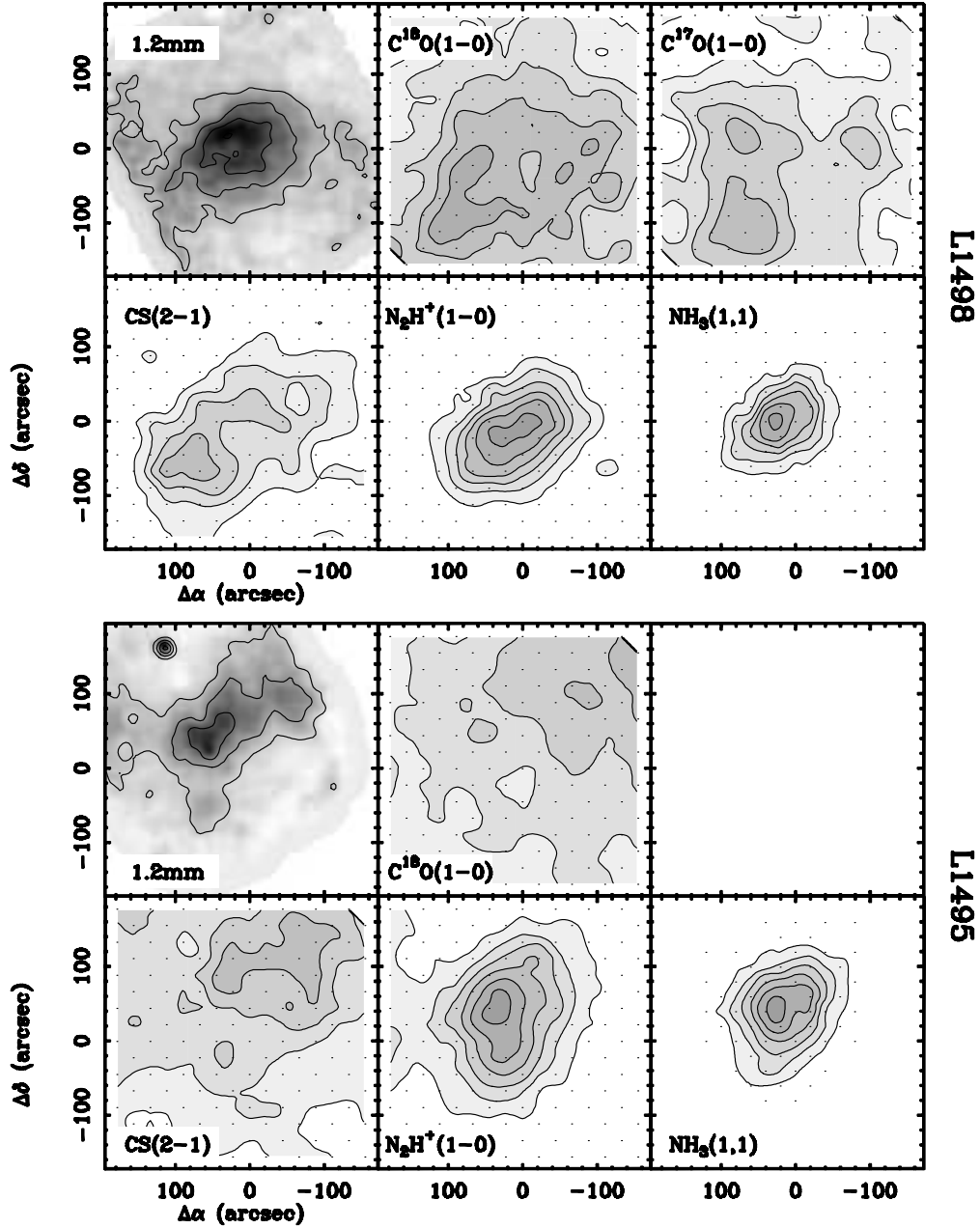}
\caption{
Maps of 1.2mm continuum (IRAM 30m telescope), C$^{18}$O(1--0), C$^{17}$O(1--0),
CS(2--1), N$_2$H$^+$(1--0) (FCRAO telescope) and NH$_3$(1,1) (100m telescope)
for L1498, L1495, L1400K, L1517B, and L1544 (L1544 1.2mm continuum map from
\citealt{war99}). Central coordinates are given in Table 1.
For each map, the first contour and the contour interval are
the same. In the 1.2mm maps of L1498, L1495, L1400K, and L1517B, contours
start at
5 mJy/11$''$-beam, although the maps have been convolved to a resolution
of $20''$, and in the L1544 map, contours start at
20 mJy/13$''$-beam. The line maps represent integrated intensities
including all hyperfine components. The first contour of
each map ordered from left to right and from top to bottom are as
follows (all in K km s$^{-1}$, in the main beam temperature scale):
L1498 (0.2, 0.075, 0.15, 0.3, 1.5); L1495 (0.3, 0.15, 0.45, 1.5);
L1400K (0.2, 0.1, 0.3, 1.0); L1517B (0.2, 0.065, 0.1, 0.3, 1.5),
L1544 (0.3, 0.13, 0.15, 0.55, 2.0). Note that the
point source to the NE of L1495
in the 1.2mm continuum map is IRAS 04112+2803 (aka CW Tau).
\label{fig1a}}
\end{figure*}


\begin{figure*}
\figurenum{1}
\plotone{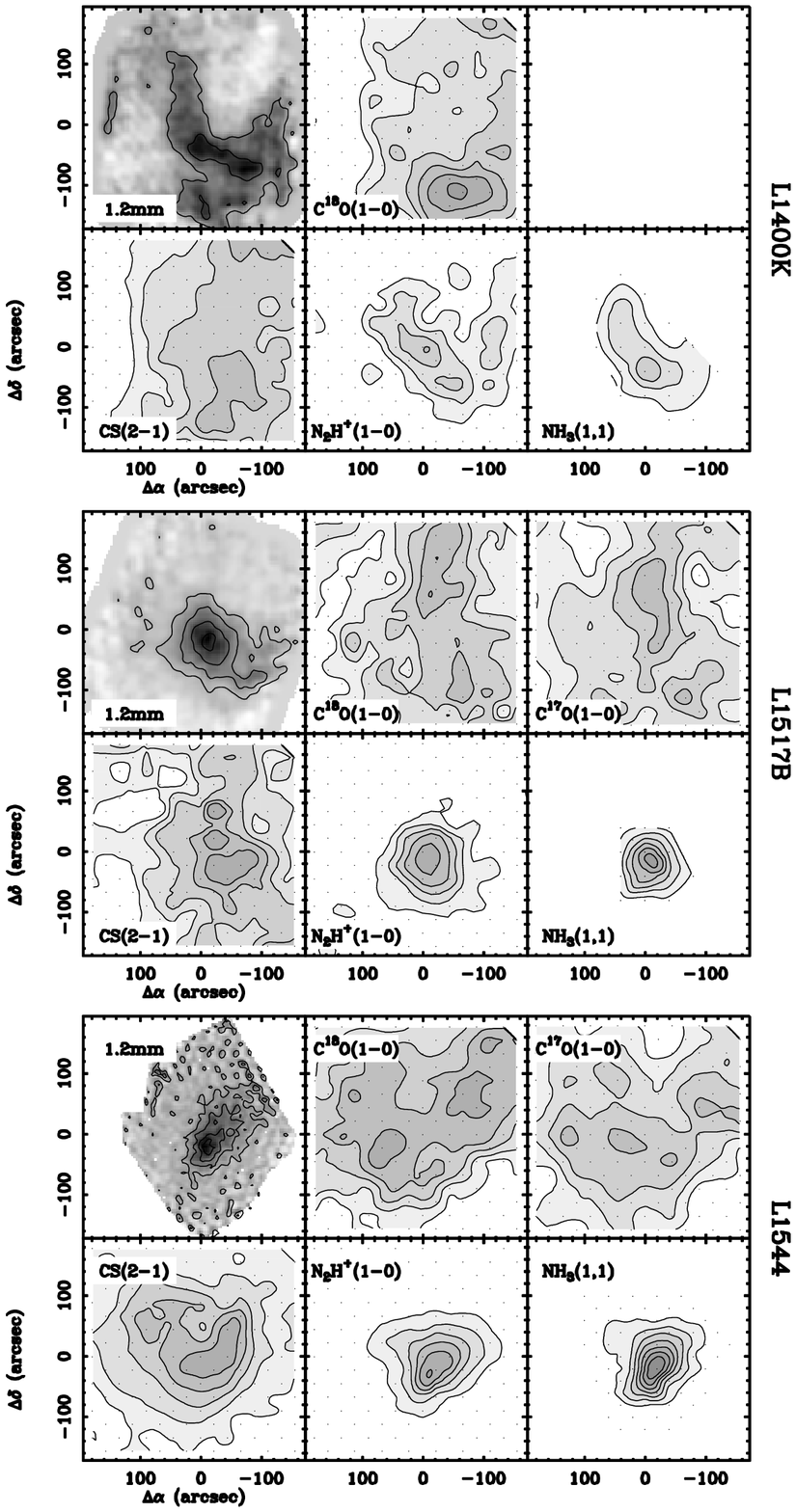}
\caption{
Continuation.
\label{fig1b}}
\end{figure*}
 

Figure 1 shows our full data set in the form of 
integrated intensity maps. For each 
starless core (L1498, L1495, L1400K, L1517B, and L1544), 
the three top panels display maps of what are expected to be 
column density tracers (dust plus the low dipole moment isotopomers
C$^{18}$O and C$^{17}$O), and the three bottom panels display maps
of high density tracers (the large dipole moment species CS, N$_2$H$^+$,
and NH$_3$). (The L1544 1.2mm map is from \citealt{war99}.) As it can
be seen by simple inspection, there is a systematic 
pattern of emission shared by all cores. On the one hand, the
continuum, N$_2$H$^+$, and NH$_3$ maps are centrally concentrated,
have approximately the same peak position and share very similar shapes 
(nearly round in L1498, L1495, and L1517B, and more elongated in 
L1400K and L1544). The C$^{18}$O, C$^{17}$O, and CS maps, on the other
hand, are much more diffuse and fragmented, and have maxima
which do not coincide with those of the dust, N$_2$H$^+$, and NH$_3$.
In fact, many maps of the CO isotopomers and CS have relative 
minima at the position where the other molecules peak, suggesting almost
an anti correlation between the two groups of tracers.

The contrast between the centrally peaked dust emission and the 
fragmented C$^{18}$O and C$^{17}$O maps (top rows in Fig. 1)
is especially striking, given that dust and C$^{18}$O/C$^{17}$O
are both expected to trace the column density of the invisible 
H$_2$ component. We can rule out a major distortion in the 
CO isotopomer maps due to optical depth or saturation effects,
since there is excellent agreement between the C$^{18}$O map and 
that of the rarer C$^{17}$O in each
core where we have observed C$^{17}$O (Fig. 1). 
In addition, the intensity ratios between C$^{18}$O and C$^{17}$O are close 
to the expected isotopic ratio of 3.65, and a hyperfine analysis of the 
C$^{17}$O multiplet shows a negligible optical depth ($< 0.1$). We
therefore conclude that both species are optically thin (see section 5.3 
for a quantitative analysis).

Temperature gradients cannot be causing 
the morphological differences either, as the relative intensities of
NH$_3$(1,1) and (2,2) indicate a constant gas temperature of about 
10~K in all observed sources (section 5.5). The cores moreover, 
being starless, lack internal heating sources to warm up the dust.
Thus, we conclude that the
difference between the continuum and C$^{18}$O maps has 
arisen from spatial differences in the dust and C$^{18}$O column densities. 
The excellent agreement between the dust emission and the emission
from the high density gas tracers N$_2$H$^+$ and NH$_3$ (Fig. 1) 
suggests that the 1.2mm continuum is a faithful tracer of the gas 
component, especially at the core center,  and that the CO isotopomers 
miss the high density central region. This must be the result of a 
drop in CO abundance at high density, similar to that
found in several starless cores by other authors 
(L1498: \citealt{wil98}, IC5146: \citealt{kra99,ber01}, L977: \citealt{alv99},
L1544: \citealt{cas99}, L1689B: \citealt{jes01})

Before estimating the abundance profile of each molecular species using
a full radiative transfer analysis, we present a simple argument to 
estimate a lower limit
to the CO abundance drop at some of the core centers. For this, we use 
the maps in Fig. 1, which show that the C$^{18}$O and C$^{17}$O 
emissions not only avoid the dust peak, but in most cases
present a relative minimum at that position. 
Simple radiative transfer shows that the intensity of an optically 
thin, thermalized line is proportional to the molecular column 
density, so at a point in the map with impact parameter $b$, the CO isotopomer 
emission is proportional to
$$\int n_{H_2}(r) X(r) dz,$$
where $n_{H_2}$ is the H$_2$ number density, $X$ is the molecular 
abundance, $z$ is the length along the line of sight, and 
$r=\sqrt{b^2+z^2}$. In order to reproduce the observed C$^{17}$O and C$^{18}$O 
emission minima toward the core peaks, the amount inside the integral 
has to reach a relative minimum, and this can only happen if the 
molecular abundance decreases to more than compensate the
central density peak.  In other words, the observed C$^{17}$O and C$^{18}$O
emission minima imply that the fraction of 
C$^{18}$O and C$^{17}$O decreases faster than $1/n(r)$ at the core centers.

\section{Core Modeling: Method Description and Density Profiles}

\subsection{Method Description and Justification}

In the following sections we present the radiative transfer modeling 
needed to derive the full chemical composition of the cores. Before
discussing all its details, we review here the main steps and 
assumptions involved in the process.

A basic simplification in our calculations is that we model the cores  
as spherical systems.
The cores were selected from previous work \citep{cas01}
for having nearly circular contour maps. The maps of
continuum, N$_2$H$^+$, and NH$_3$ emission in Fig. 1 show that this is in
fact the case in most systems (L1400K being the most deviant core).
Although some cores would seem better modeled as spheroids, our
lack of information on the line-of-sight component makes any guess about
this third dimension highly uncertain. It seems, therefore, more convenient
to use spherical models for the cores and to compare the predictions of
those models with azimuthal averages derived from the data. This will
be our approach in what follows.

The first step in our modeling is to derive density and temperature 
distributions for each core. We determine the density using
the 1.2mm continuum emission, which seems 
a reliable tracer \citep[e.g.,][]{and96, war99}
and for which we have the data of highest angular resolution. To measure 
the gas temperature, we have carried out a standard LTE analysis 
of the NH$_3$ data and derived for each core a constant temperature of
approximately 9.5~K, except for L1544 for which we derive 8.75 K. These 
values are in agreement with other estimates of these and
similar cores \citep[e.g.,][]{ful93}, and they will be later confirmed 
by our full, non LTE NH$_3$ analysis in section 5.5. 

	Once densities and temperatures have been estimated, we derive 
molecular abundance profiles by solving the line radiative transfer 
with a slightly modified version of the Monte Carlo 
code from \citet{ber79}. For each core we vary the molecular abundance 
of each species and the gas velocity field until the model emission 
(convolved with the appropriate Gaussian beam) fits both 
the observed radial 
distribution of integrated intensity and the shape of the emerging
spectrum toward the core center. This approach is as close as we can 
get from modeling the emission at each core position 
with our spherical models \citep[see, e.g.,][for a similar procedure]{bie93}.

	As a result of the above modeling, we derive a set of physical 
and chemical parameters that, for each core, simultaneously fit all the 
data in Fig. 1 plus the emergent central spectra. 
Fitting such a large number of observational data simultaneously imposes 
strict constrains in the model parameters,
and this has made their determination a highly non linear process.
Although the description in the
following sections is sequential, the real fitting process
required a number of iterations and back-and-forth
modeling until self-consistency was achieved.

\subsection{Density Profiles}

We start our analysis by deriving a density distribution for each core, 
and we do this by modeling its dust 1.2mm emission. As our
models assume spherical symmetry, we first create 
spherical equivalents of each core by taking azimuthal averages of the data.
For the roundest cores L1517B and L1495, we simply average the data in 
circles, 
and for the more elongated L1498, L1400K, and L1544, we average the emission
along ellipses with aspect ratio ($b/a$) and position angle ($PA$) as
indicated in Table 2. In these latter cases, the radial coordinate
is the geometric mean of the major and minor axes of the averaging ellipse.
The resulting profiles, both in linear and logarithmic scale, are presented in
Figure 2.


\begin{figure*}
\figurenum{2}
\plotone{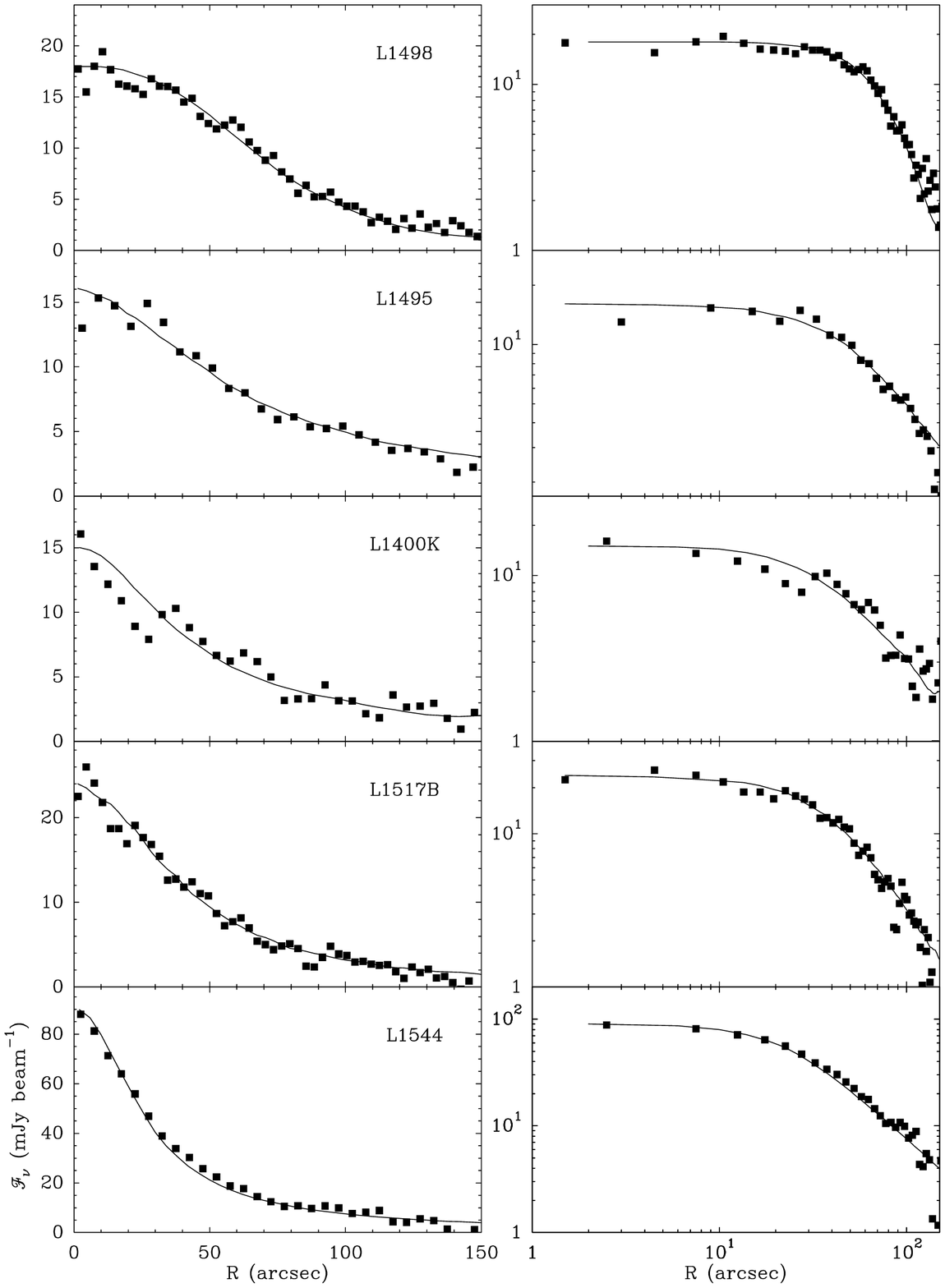}
\caption{
Radial profiles of 1.2mm continuum emission in both linear and log-log scale.
(L1544 data from \citealt{war99}.)
Solid squares are telescope measurements and lines are models for which
a simulation of an on-the-fly observation has been done (see text).
For L1495 and L1517B, the profiles represent circular averages, while
for L1498, L1400K, and L1544, they represent elliptical averages with
the parameters shown in Table 2. See also Table 2 for the input parameters
of the models. Scale is in mJy/11$''$-beam for L1498, L1495, L1400K,
and L1517B, and mJy/13$''$-beam for L1544. In contrast to Fig. 1, the
data from the first four cores have not been convolved additionally after
the telescope observation.
\label{fig2}}
\end{figure*}


The goal of our modeling is to find a set of density distributions 
whose 1.2mm dust emission fits the radial profiles shown in Fig. 2,
and this requires first choosing a set of physical parameters for the dust 
grains. Following \cite{and96}, we use a 
1.2mm emissivity $\kappa_{1.2 mm}= 0.005$ cm$^2$ g$^{-1}$, although it should
be noted that this parameter has an uncertainty of about a factor of 2.
For dust temperature, we use a constant value equal to the
gas temperature estimated from the NH$_3$ analysis (Sec. 5.5).
Recent work by \citet{eva01} and \citet{zuc01} 
suggests that the dust temperature in a core may decrease toward the center
due to 
extinction in the warming interstellar radiation field. Our NH$_3$ data,
however, suggests that the gas temperature is constant across the core
(Sec. 5.5), and given that the central densities we derive
($\sim 10^5$~cm$^{-3}$) are high enough for gas and dust to be thermally
coupled \citep{gol78, gol00}, we conclude that if central cooling
is present, it occurs at scales smaller than the resolution of our
molecular line data, so it can be safely ignored in our models. 

If the dust kinetic temperature and the dust emissivity are constant,
the emitted flux density from the core at the bolometer central
frequency (240 GHz) is simply given by
\begin{eqnarray*}
S_{1.2mm} & = & \Omega_{beam}\; \kappa_{1.2mm}\;  m\; N(H_2)\; B_\nu(T_d)\\
& = & 5.2 \times 10^{-11}\; B_\nu(T_d)\; [\theta('')]^2\; N(H_2)\;\; 
{\rm mJy/beam},
\end{eqnarray*}
where $\Omega_{beam}$ is the telescope
beam solid angle, $m$ is the mean molecular
mass, $N(H_2)$ is the H$_2$ column density (in cm$^{-2}$ in the lower
equation), $\theta('')$ is the FWHM of the beam in arcsec, and 
$B_\nu(T_d)$ is the Planck function at the dust temperature $T_d$.
With this equation, finding the
core density law becomes simplified to finding the function $n(r)$ that gives
rise to the appropriate $N(H_2)$ profile. Unfortunately,
the process is complicated by the need
to take into account the  beam smoothing and spatial filtering introduced
by the on-the-fly bolometer observation. This
problem, discussed in detail by \citet{mot01}, requires
that a simulation of the  on-the-fly observation is carried out
before comparison with the data. In our case, we have used
the NIC software, which lets us simulate an on-the-fly observation
of the model with the same parameters as those used during the
telescope run, and reduce the simulated data the same way the
real data were reduced. This guarantees that data and model
are properly compared.

To find the density distributions that reproduce the observed
radial profiles, we iterated the fitting procedures until
reasonable convergence. Previous work
has shown that single power-law density profiles do not fit the emission
from starless cores, and that 
a central flattening is always needed to reproduce the data 
\citep{war94,and96,bac00,alv01}. Thus, it has become a standard practice to use 
double power laws with the inner portion nearly flat. The 
discontinuous derivative of these fits, however, seems rather artificial,
and it appears more likely that real cores
have smoother density profiles, with 
the double power-law expressions being just an approximation.
For this reason, we have searched for a family of analytic 
density profiles that combine the
power-law behavior for large $r$ and the central flattening at small $r$.
After different tests, we have chosen profiles of the form
$$n(r) = {n_0\over 1+(r/r_0)^\alpha}, $$
where $n_0$ is the central density, $r_0$ is the radius of the ``flat'' region 
($2 r_0$ is the full width at half maximum), and $\alpha$ is 
the asymptotic power index. As Fig. 2 shows,
this family of curves, with the values for $n_0$, $r_0$, and  $\alpha$
indicated in Table 2, fits each of the observed radial brightness
distributions with great accuracy, and it therefore constitutes the
basis of our core modeling. Values of $n_0$ range between about
$10^5$ and $10^6$ cm$^{-3}$, and their major source of uncertainty is their
linear dependence on $\kappa_{1.2mm}$ (which has a factor of 2 
uncertainty). $r_0$ is probably the best 
determined parameter 
and ranges from 3,000 to 10,000 AU, with an uncertainty of
about 1,000 AU. Finally, the $\alpha$ parameter ranges from 2 to
4, and is probably accurate to 0.25-0.5.
(\citet{whi01} have recently proposed a similar
expression, although they require $\alpha \ge 4$, which seems outside
the range of our empirically determined values. See also \citet{lan01} for
an alternative fit to the L1498 density profile.)

\section{Abundance Profiles: Evidence for Chemical Differentiation}

\subsection{Monte Carlo Model Parameters}

Once the core density profiles have been determined from the continuum 
emission, we derive molecular abundances for all the observed 
species by fitting their distribution of  
line intensities. We do this by solving the equations of radiative transfer
with a Monte Carlo code \citep{ber79}, and as mentioned before, we
assume a gas kinetic temperature of 8.75 K for L1544 and 9.5 K for the 
rest of the cores (section 5.5 shows how these 
temperatures are required by the NH$_3$ data). 

For each core, the free parameters in our Monte Carlo models are the 
molecular abundances, the gas velocity field (both
systematic and turbulent), and the core maximum radius (almost immaterial
given the steep density profiles). The velocity field
is well constrained by the need to  fit simultaneously the central 
line shape of C$^{18}$O(1--0), CS(2--1), N$_2$H$^+$(1--0), NH$_3$(1,1),
and NH$_3$(2,2).
In this way we find that all cores but L1544 are consistent with a constant 
turbulent linewidth, with a value of FWHM = 0.167 km s$^{-1}$ (variable 
$vturb=0.1$~km s$^{-1}$ in the \citealt{ber79} Monte Carlo program).
For  L1544, we model the emission with a turbulent linewidth
of 0.13 km s$^{-1}$ at the outermost radius increasing inward
as $r^{-0.5}$ up to a value of 0.25 km s$^{-1}$ (\citealt{taf98} also
found an inward increase of the turbulent linewidth, and recently
\citealt{cas01c} have found a similar pattern studying the linewdith
in the plane of the sky). This larger
value of the linewidth in L1544 probably reflects the complicated
kinematics of this core, already discussed by \citet{taf98} and
 \citet{cas01c}.

Together with the turbulence velocity field, we find
that some cores have systematic velocity gradients along the line of
sight. These gradients are required to fit the displacement of
the CS self absorptions with respect to the systemic velocity (section 5.2),
and, for simplicity, we parametrize the velocity field with a linear
function. The values we find for $dV/dr$ (Table 3) are consistent with
the typical core velocity gradients across the line of sight measured
by other authors (e.g., \citealt{goo93}).

The maximum core radius, as mentioned before, is not a critical
parameter in our modeling, because the steep density gradient
of each core sets an almost natural outer edge.
All cores but L1544 are well fit with a maximum
core radius of $190'$ ($4\times 10^{17}$ cm at the distance of Taurus and
$5\times 10^{17}$ cm for L1400K), and L1544
requires twice that value. This large radius for L1544 is needed in order to 
fit its very deep CS self absorption, a result already found by \citet{taf98}.
These authors carried out a similar Monte Carlo fitting of the CS lines
in L1544, but ignored possible abundance 
variations with radius. The models presented in this paper supersede
those in \citet{taf98}.

A number of tests have been carried out to ensure that the Monte Carlo results
presented here are independent of the internal program parameters.
During searches for the best fits, the core was divided into 100
shells spaced logarithmically, the number of model photons was set 
to 2000, 40 iterations were performed, and no reference radiation
field was used. Previous experience had shown that these parameters
were sufficient to avoid dependence on the initial conditions and to
reach a stable solution. As a further test, all best models for L1517B 
(one of the most regular cores) have been repeated doubling the number of
shells and iterations and increasing the number of photons by an 
order of magnitude; no significant difference has been found between these
new runs and the standard cases. We take this stability of the results
as an indication that the 
outputs from our models are true solutions to the molecular radiative
transfer problem, and are
not affected by the internal details of the Monte Carlo code.

\subsection{C$^{18}$O and CS Abundances}

C$^{18}$O and CS are closed-shell linear rotors with 
known collisional rates, and this makes their radiative transfer
the easiest to calculate among all our species. For C$^{18}$O, we use
the molecular constants from \citet{lov74} together with 
the rate coefficients of collisional excitation by para H$_2$ from 
\citet{flo85}.
For CS, we also use the molecular constants from \citet{lov74}, 
together with the collisional coefficients 
for para H$_2$ from \citet{gre78}. 
A total of 9 rotational levels are considered for each molecule,
and the resulting model line intensities 
are convolved with a Gaussian beam to simulate the result of an 
observation with the FCRAO telescope (FWHM of $46''$ for C$^{18}$O(1--0)
and $50''$ for CS(2--1), according to \citealt{lad96}).

Our first set of models assumes spatially constant abundances for 
both C$^{18}$O and CS. For C$^{18}$O, we assume an abundance 
of $1.3 \times 10^{-7}$ for all cores
and for CS we use different abundances so that we
approximately fit the emission at large radius ($>100''$) for each core. 
The resulting radial profiles, indicated by dashed lines in the left panels 
of Fig. 3, fail to reproduce the observed intensities near the core center
by more than a factor of 2 and are, therefore, inconsistent with the data.
A similar set of constant abundance models
with an abundance low enough to fit the central emission
(models not shown in Fig. 3) also clearly fails to fit the 
C$^{18}$O and CS radial profiles, in this case by underestimating the
emission at large radius.

In order to improve the fits, we have run additional 
models with a central abundance drop.
In section 3 we have seen that the C$^{18}$O abundance has to decrease
faster than $1/n(r)$, so with the Monte Carlo program, we have explored 
different analytical expressions of increasing steepness:
$1/n(r)^2$, an exponential form, and a central abundance hole. 
The results derived from these models are practicably indistinguishable
due to the limited resolution of the FCRAO data ($\sim 50''$), so, for
simplicity, we will only present here the results derived with 
the exponential abundance law:
$$X(r) = X_0 \exp (-n(r)/n_d),$$
where the free parameters $X_0$ and $n_d$ represent the
low-density abundance limit and the e-folding size of the abundance drop.
A  more detailed discussion of the shape of
the abundance profiles using higher angular resolution data and multiple
transitions will be presented in a forthcoming paper \citep{taf01}.

\begin{figure*}
\figurenum{3a}
\plotone{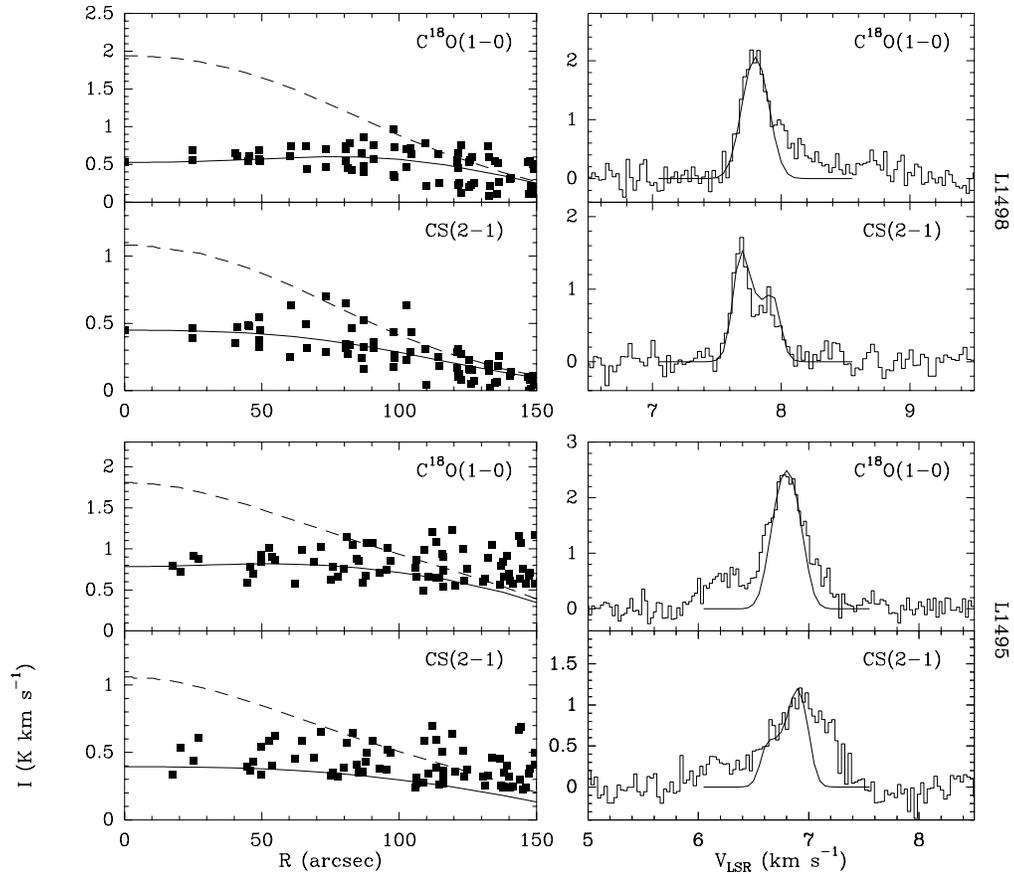}
\caption{
Radial profiles of integrated emission (left) and central emerging
spectrum (right)
for C$^{18}$O(1--0) and CS(2--1) in each of our cores. Observations
are indicated by solid squares in the radial profiles and by histograms in
the spectra (all temperature units are in the main beam scale).
The results from two radiative transfer models are plotted
over the radial profiles: the dashed lines are constant abundance models
and the solid lines are models with a rapid decrease in the central
abundance.
(Constant abundance models assume X(C$^{18}$O) = 1.3 $10^{-7}$ in all cores
and X(CS) = 3.0 $10^{-9}$ in L1495 and L1517B, 2.0 $10^{-9}$
in L1498 and L1544, and 6.0 $10^{-9}$ in L1400K. See Table 4 for parameters
of the models with a central abundance drop.)
Note how the constant abundance models fail to fit the
central emission, while the models with variable abundance fit the
data at all radii satisfactorily.
The predicted emerging intensities of these variable abundance models
(after convolution) are presented by solid lines
overlaid on the central spectra in the right panels.
Note the presence of additional velocity components in the C$^{18}$O
emission toward L1498 and in C$^{18}$O and CS toward L1495. These
components probably arise from unrelated ambient gas and
have not been included in the integrated emission of the radial profiles.
\label{fig3a}}
\end{figure*}

\begin{figure*}
\figurenum{3b}
\plotone{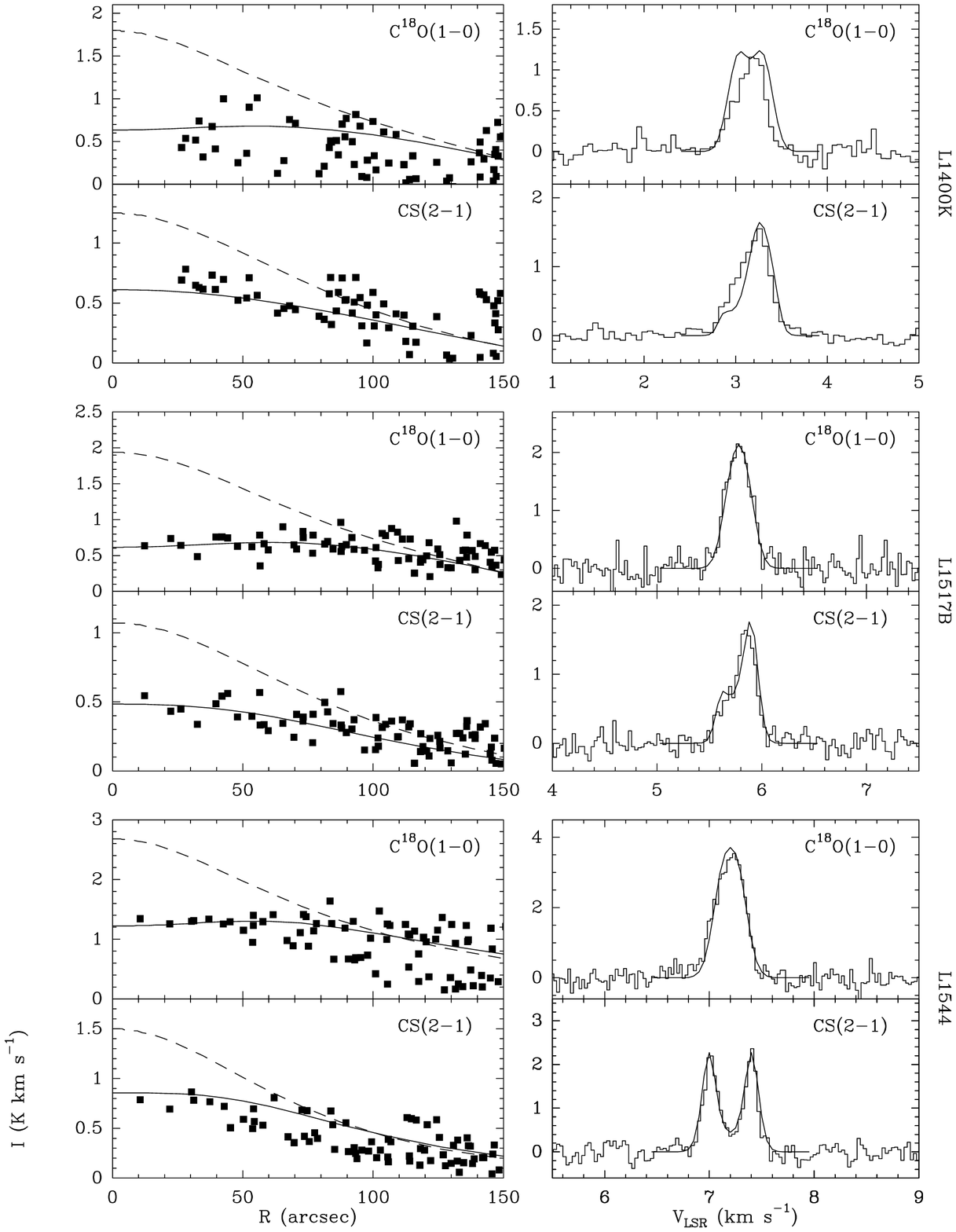}
\caption{
Continuation.
\label{fig3b}}
\end{figure*}


The results from this last set of models are indicated by solid lines in 
Fig. 3. For C$^{18}$O, all models have the same $X_0$ value
of $1.7 \times 10^{-7}$, following \citet{fre82},  and the
$n_d$ parameter ranges from about $1.5 \times 10^4$ to 
$5.5 \times 10^4$ cm$^{-3}$, depending on the core  (see Table 4). 
For CS, different cores need slightly different values of $X_0$, but most 
estimates are close to $4 \times 10^{-9}$ (Table 4). 
The $n_d$ values for CS are also of the order of a
few  $10^4$ cm$^{-3}$ ($10^5$ cm$^{-3}$ in L1544).

As seen in Fig. 3, the exponential-law abundance models
fit the observed radial  profiles better than the constant abundance
models, especially in the inner 100$''$. For L1498 and L1517B, 
the fits are good at all radii, and the same could be said for L1544, but 
the scatter in the radial profiles makes it difficult
to choose the best fit at large radius. For L1495 and L1400K, however, 
the fits deviate from the data, and 
the scatter of the radial profiles makes it impossible to 
find a perfect fit even if we could draw an arbitrary radial profile.
The origin of these fitting problems can be understood by looking at
the maps in Fig. 1, which show that, in these two cores, the
C$^{18}$O and CS emission is so different from that
of the dust, NH$_3$, and N$_2$H$^+$ that it seems to
trace additional structures unrelated to the cores. These structures 
most likely arise from nearby lower density gas, and their modeling
is outside the reach of our simple scheme,
which uses the continuum emission as the basis of the core model
and assumes spherical symmetry. This last assumption, in addition,
has the most extreme deviation in the case of L1400K, which is
clearly elongated NE-SW and has additional extensition to the
west (Fig. 1). The above limitations in the L1495 and L1400K models, however,
do not contradict the idea of a C$^{18}$O and CS abundance decrease
in these two cores.
Indeed, it seems that their central abundance decrease
is more severe than that shown by other cores, and 
it is that very fact which makes them almost
invisible in C$^{18}$O and CS.

An additional requirement for our models is that they also 
fit the central spectrum, both in integrated intensity and detailed 
shape. As the right panels of Fig. 3 show, this is the case for
L1498, L1517B, and L1544, while there are differences between model and 
data in L1400K and L1495 (note that the extra wing in the L1498 C$^{18}$O
spectrum most likely arises from the extra
red component seen in the maps of \citealt{lem95}). As 
with the radial profiles, the worst fits occured in the cases of 
L1495 and L1400K. This 
probably results from the difficulty in modeling extended, non
spherically-symmetric material,
a situation further complicated in L1495 by the presence of gas at 
different velocities (see our C$^{18}$O spectrum and \citealt{goo98} for
a detailed discussion of the velocity structure in L1495). 
For the cores with good fits
(L1498, L1517B, and L1544), the agreement between model and data
is achieved by finding the appropriate combination of 
systematic and turbulent velocity fields, a process carried out
simultaneously with the modeling of the N$_2$H$^+$ and NH$_3$ 
emission (sections 5.4 and 5.5) because our model requires that, 
for each core, all lines be reproduced with a single velocity field.

The CS spectra are particularly sensitive to the systematic velocity field
due to the presence of self absorptions. These features naturally 
arise from the combination of a drop in excitation caused by the density
fall off with radius and a relatively large abundance of CS 
in the outer core layers. Regarding the modeling of
the emission, the difference in velocity between the self absorption 
dip and the line center constrains any possible velocity 
gradient along the line of sight. Most of our cores show
evidence of such gradients, which we have modeled for simplicity 
using linear functions with the coefficients given in Table 3.
We find both inward and outward velocity fields among
our sample cores, with the best examples of each case being L1498 (inward) and 
L1517b (outward). We note that our model for the infall case 
L1544 lacks a gradient
along the line of sight. This occurs because our modeling process was 
restricted to fit only the shape of the central spectrum, which appears rather 
symmetric and deeply self-absorbed (see Fig. 3). As \citet{taf98}
have shown, L1544 presents evidence of extended inward motions
in CS when off-center positions are considered, an effect that we cannot
model here because of our assumption of spherical symmetry.

To conclude our analysis of the  C$^{18}$O and CS emission, we will
estimate the abundance contrast between the core center
and the outer layers implied by our models. 
To do this, we define for each core the parameter
$f$ as the ratio between the model
abundances at $r=20''$ (half our beam) and $r=100''$. This $f$ parameter,
shown in Table 4, is typically of the
order of $10^{-2}$ for both C$^{18}$O and CS, indicating abundance
drops for these two molecules of two orders of magnitude in each core.
Such strong abundance drops are, in
fact, consistent with a case in which the cores have no C$^{18}$O and CS 
molecules at their centers.

\subsection{C$^{17}$O and C$^{34}$S Data}  

The Monte Carlo modeling takes 
into account optical depth and trapping effects, so the 
poor fit of the constant abundance models in Fig. 3
cannot be explained by a failure to take those factors into accout.
Still, the C$^{18}$O(1--0) lines have moderate optical depth 
and the CS(2--1) spectra are obviously thick and self absorbed, so
it seems appropriate to test the results from the previous section using 
optically thin lines of C$^{17}$O and C$^{34}$S.
Here we carry out such testing with
C$^{17}$O(1--0) observations of the cores with brightest C$^{18}$O
emission (L1498, L1517B, and L1544, see Fig. 1) and 
C$^{34}$S(2--1) data from the most self absorbed core (L1544).

The C$^{17}$O molecule has hyperfine structure, and this in principle could
complicate its radiative transfer modeling. The 
low optical depth and dipole moment of this molecule, however, causes
its level excitation to be dominated by collisions, leaving
trapping and splitting effects negligible. This allows us
to model the C$^{17}$O radiative transfer with the Monte Carlo 
code as if there were no hyperfine structure, and to predict the
combined intensity of the three hyperfine components. The lack of
model prediction for the individual components is of little consequence,
as the low signal-to-noise ratio of our C$^{17}$O data
makes it necessary to combine the observed components 
before comparing them with the model predictions (see sections 5.4 
and 5.5 for 
model line predictions for species with hyperfine structure like 
N$_2$H$^+$ and NH$_3$).

To test the C$^{18}$O calculations from the previous section, we 
need to create equivalent models for the rarer C$^{17}$O. As the relative 
abundance of C$^{18}$O over C$^{17}$O seems constant over
the Galaxy with a well measured value of 3.65 \citep{pen81},
the C$^{18}$O abundance curves determined previously can be automatically 
converted into C$^{17}$O abundance profiles just by dividing by 3.65.
This means that our C$^{17}$O Monte Carlo modeling is a zero-free-parameter
calculation, as it has no internal values that can be used to 
improve the fitting. 


\begin{figure*}
\figurenum{4}
\epsscale{1.0}
\plotone{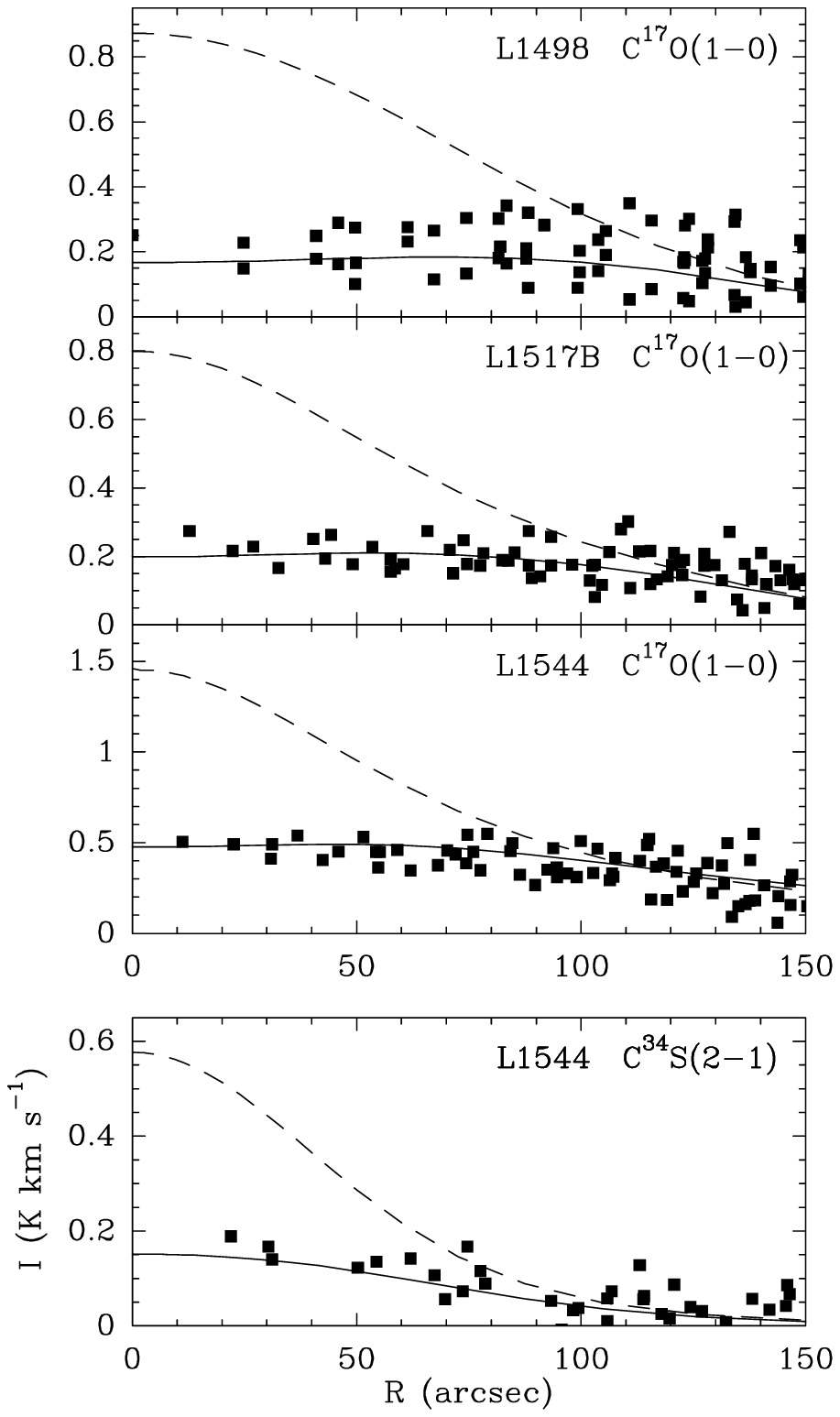}
\caption{
Radial profiles of integrated emission
for C$^{17}$O(1--0) (three top panels) and C$^{34}$S(2--1) (bottom panel)
toward selected cores.
Observations are indicated by solid squares and results
from radiative transfer models are indicated by lines. The models
assume the same abundance profiles
as those in Fig. 3, with the ISM isotopic ratio of
3.65 for C$^{18}$O/ C$^{17}$O \citep{pen81} and the Solar
ratio of 22.7 for C$^{32}$S/C$^{34}$S.
Note how the constant abundance models (dashed lines)
miss the data toward the core centers by large factors. The models with
central abundance drops (solid lines), in contrast, automatically fit the
data without adjusting any additional parameter.
\label{fig4}}
\end{figure*}


The top panels of Fig. 4 present a
comparison between the results of our C$^{17}$O(1--0) models
and the FCRAO observations. Here, once again, the dashed lines represent 
constant
abundance models and the solid lines indicate models with an exponential 
abundance drop. As before, the models with constant abundance fit only the
outer core emission and deviate dramatically from the observed intensity
toward the center. The discrepancy between the C$^{17}$O constant abundance 
models and data is larger than seen for C$^{18}$O,
due to the lower optical depth of the rarer isotope, and
reaches values of up to a factor of 5 toward the core centers.

In contrast to the poor fit of the constant abundance models, 
a remarkable fit is obtained with the models that have
an exponential abundance drop (Fig. 4). These  models have 
the same abundance profiles as those
used for C$^{18}$O (Table 4), with the only 
difference being that the $X_0$ value has been
divided by 3.65. The good match of these models, 
both at large and small radii, constitutes a further and final
proof that the abundance of the CO isotopes decreases sharply and 
systematically at high densities in all cores of our sample.

Our modeling of the C$^{34}$S(2--1) emission is also a 
zero-free-parameter calculation. The 
C$^{32}$S/C$^{34}$S isotopic ratio in the interstellar medium equals, 
to a good approximation, the Solar value of 22.7 \citep{luc98}, so 
the C$^{34}$S abundance profiles
are just 22.7 times smaller than the CS abundance profiles. 
To test the main isotope CS 
calculations in L1544, we now run Monte Carlo radiative transfer
models for C$^{34}$S(2--1) for the cases of constant 
abundance and exponential drop.

As Fig. 4 shows, the model with constant abundance (dashed lines) is 
again inconsistent with the data, as it overpredicts
the central intensity by more than a factor of 3. The model with an
exponential abundance drop, on the other hand, again fits the data remarkably 
well, predicting a radial profile of the correct intensity and shape.
This confirms our conclusion that the CS abundance decreases
rapidly toward the core center and shows that our Monte Carlo modeling
can predict accurate abundance profiles even in a case with
large optical depth and strong self absorption.

\subsection{N$_2$H$^+$ Data}  

Solving the radiative transfer for N$_2$H$^+$ is much more complicated than 
for CO and CS, as the two N atoms of N$_2$H$^+$ make its rotational 
level structure split into multiple hyperfine components
(7 for J=1--0, 38 for 2--1, 45 for 3--2, etc.).
The large number of resulting transitions, and their possible overlap,
makes the solution of the radiative transfer a formidable task,
and places  it outside the scope of our present 
analysis. In this section, we discuss how to simplify the N$_2$H$^+$
radiative transfer treatment, and how to derive reasonably-accurate 
molecular abundance profiles for all our cores.

	The presence of hyperfine (hf) splitting introduces two 
new elements into the radiative transfer.
First, the splitting allows the relative population of the 
hyperfine sublevels to depart from their statistical weights. 
This effect, which gives 
rise to hyperfine multiplets with ``non-LTE'' intensity ratios, has
been observed in N$_2$H$^+$(1--0) toward starless cores by \citet{cas95}.
Although well-detected, this effect involves less than 10\%
of the emitted 1--0 radiation \citep{cas95}, so it must 
represent only a small perturbation in the total emerging flux. 
In the simplified treatment presented here, we will ignore
this effect assuming that inside each rotational transition the 
sublevels are populated according to their statistical weights.

	The second effect of the hyperfine splitting is a
change in the line trapping. Because of the splitting, the 
photons from each rotational transition can escape more easily, and this
decreases their role in excitation. This effect will 
only be significant if trapping dominates collisions as the main
excitation mechanism, and this can only occur at very large optical depths. 
A simple LTE fit to the observed (1--0) multiplet (the one expected to
be thickest) shows that, in our cores, the averaged optical depth of the
hf components does not exceed 2 even at the
core maximum, and that in most cases the lines are optically thin. This 
suggests that collisions dominate the excitation, and 
that the trapping decrease represents only a minor
contribution. To further explore this point, we have run radiative
transfer models 
using two extreme limits. In one, we completely ignore the hf splitting, 
so the trapping is maximum, and in the other, we artificially turn off 
the trapping so it has no effect on excitation. The real
case will necessarily lie between these two
limits, closer to one or the other depending on optical depth. 
As a result of this test, we find that for the conditions of our cores,
both limits predict the same shape of the radial profile of integrated 
intensity. Thus, the shape of the N$_2$H$^+$ abundance curve we derive 
from fitting the radial profile is independent of our treatment of the hf 
splitting. The absolute value of the abundance, in addition,
typically differs between the two cases by just 30\%
(80\% in L1495), indicating that the effect of the treatment of the 
hf splitting is small, and probably comparable to the uncertainties 
in the collisional coefficients (see below).

Given the small effect of the hf splitting in the excitation, we can 
simplify the radiative transfer calculation by separating the problem in two
steps: the solution of the level excitation and the 
prediction of the emergent spectrum. In the first step,
we use the Monte Carlo code without hf splitting, and calculate the
combined population of each rotational level. For the collisional 
rate coefficients, we use the recent set for collisions between HCO$^+$
and para H$_2$ from \citet{flo99}. As \citet{mon84} has shown, the similar
rotational constants and electronic structures of N$_2$H$^+$ and HCO$^+$ 
imply that the two species have similar cross sections, and they do
in fact have collisional rates with He which are indistinguishable
\citep[see also][]{gre75}. Until a specific calculation for collisions
between  N$_2$H$^+$ and H$_2$ exist, the use of the  HCO$^+$ rates
remains the most accurate choice.

Once the combined population of the rotational levels has been derived
with the Monte Carlo code, we take into account the full hf structure to
integrate the equation of radiative transfer along the line of sight 
and predict the emergent 1--0 spectrum. We 
assume that the hf sublevels are populated according to their 
statistical weights and use the relative line frequencies from 
\citet{cas95}. This emerging spectrum is finally convolved with a Gaussian 
function of 52.5 arcsec FWHM, to simulate the smoothing effect of the FCRAO 
telescope beam.

\begin{figure*}
\epsscale{1.3}
\figurenum{5}
\plotone{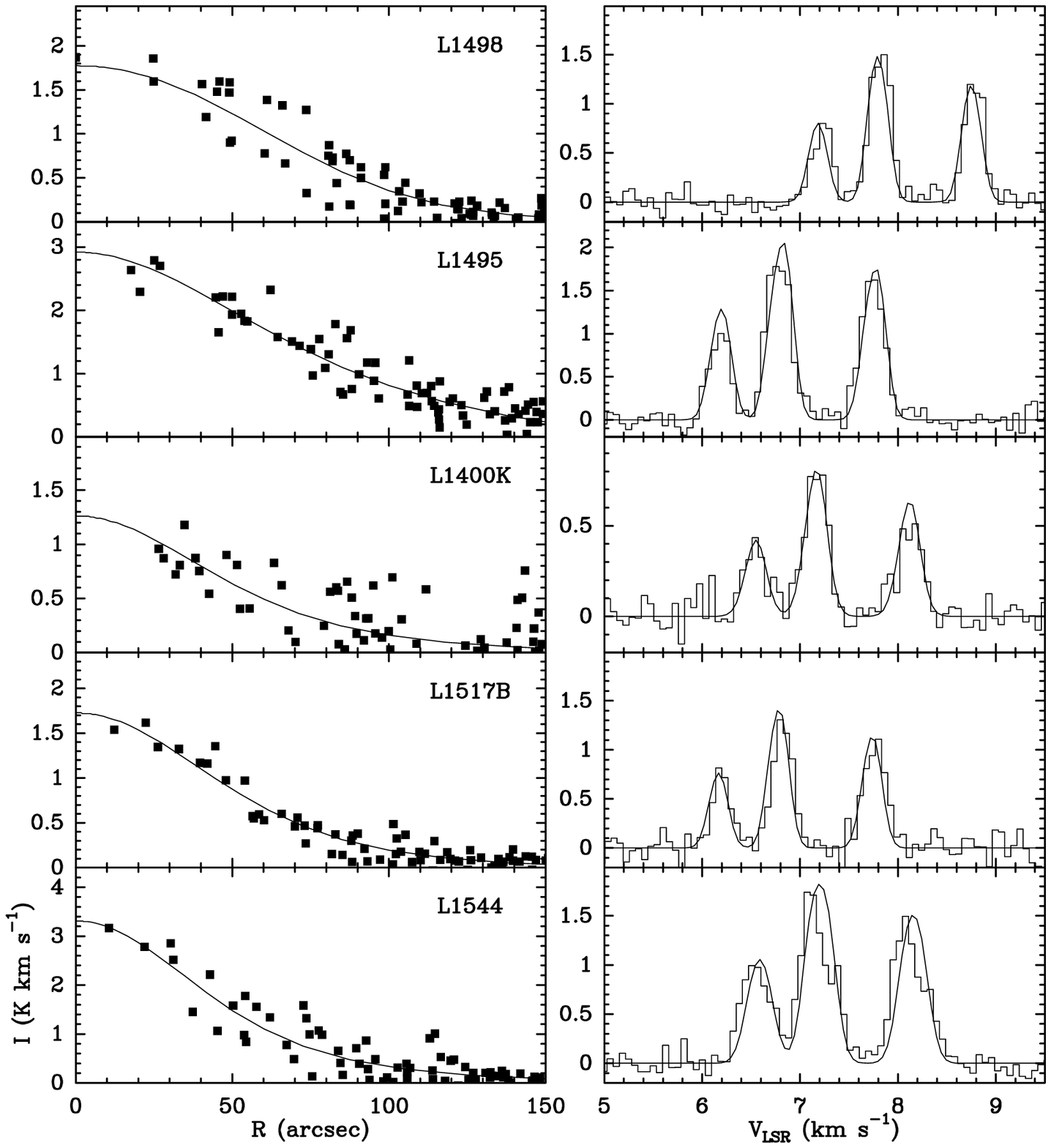}
\caption{
Radial profiles of integrated emission (left) and central emerging
spectrum (right) for N$_2$H$^+$(1--0). Observations
are indicated by solid squares in the radial profiles (sum of all hyperfine
components) and by histograms in the spectra (all temperature units are in
the main beam scale).
Constant abundance models are indicated by solid lines, and
unlike the case of C$^{18}$O and CS (Fig. 3), they accurately
fit the data at all radii (see parameters in Table 4).
For the emerging spectra,
only the central hyperfine group is presented to show in detail the
good matching of both the linewidth and the hyperfine ratios.
To use the same velocity scale in all the spectra, both data and
model for L1400K have been shifted by 4 km s$^{-1}$, and by 1 km s$^{-1}$
for L1517B.
\label{fig5}}
\end{figure*}


The results of the N$_2$H$^+$ calculations are presented in Fig. 5, where
constant abundance models are compared with FCRAO observations for
each core in our sample. As Fig. 5 shows, models with constant 
N$_2$H$^+$ abundance provide satisfactory fits in most cases 
to both the radial profile
of integrated intensity (left panels) and the central emerging
spectrum (right panels), in sharp contrast to what was found 
for C$^{18}$O and CS. Although there are small deviations 
between model and data in Fig. 5,
they most likely result from our simplified modeling and do not 
represent real abundance 
variations across the cores. This seems to be the case in L1400K, where 
the bright emission at large radius (and overal
larger scatter) is due to the elongated shape of this core and to
the presence of an extra N$_2$H$^+$ component to the west (see
Fig. 1). Also, our inability to fit the non-gaussian shape of the L1544
spectrum is due to the presence of two velocity components 
in this core \citep{taf98,cas01c}, which are simply modeled here with a 
broader gaussian component. Taking into account these effects (and
the scatter in the radial profiles), we conclude that our data are
consistent with all cores having constant N$_2$H$^+$ abundance, 
which we estimate to be of the order of $10^{-10}$ (see Table 5 for the value
derived for each core).

Although proving that the N$_2$H$^+$ abundance is constant has required
a full radiative transfer calculation, 
it is clear from Fig. 1 that the similarity between the 1.2mm 
continuum and the N$_2$H$^+$ data implies 
an abundance behavior of this molecule different from that 
of CO and CS. \citet{cas99} and \citet{ber01} have reached similar
conclusions for the cases of L1544 and IC5146, respectively, using
a simpler (standard LTE) analysis than the one presented here.
The different behavior of N$_2$H$^+$, on the one hand, and C$^{18}$O and CS,
on the other, is the first element of a chemical differentiation
pattern that affects all our cores, and a clear sign that
the common assumption that different molecular
species trace the core interior equally needs serious revision. 

\subsection{NH$_3$ Data}

The NH$_3$ molecule is a symmetric top with inversion doubling
and consists of ortho and para species, which coexist almost independently 
of each other (see, e.g., \citealt{ho83}). The (J,K)=(1,1) and (2,2) 
inversion lines we have observed 
arise from para-NH$_3$, so our modeling will only deal with this
species. 

Like N$_2$H$^+$, NH$_3$ has hyperfine splitting, thus, 
our treatment of this effect again involves some simplifications. Fortunately, 
the effect of the hf splitting in NH$_3$ is even less critical than in 
N$_2$H$^+$. On the one hand, NH$_3$ does not present hf anomalies toward 
dark clouds (\citealt{stu84}, also our own data), so 
the population of the hf sublevels should be proportional
to their statistical weights. On the other hand, the decrease of trapping due 
to the hf splitting has very little effect on the collisionally dominated
NH$_3$ excitation. This once again, can be seen by comparing models with 
the two extreme trapping treatments (see section 5.4), which for
L1517B (the roundest and second brightest core in NH$_3$) give
rise to an 
abundance difference of less than 10\%. As before, such a difference is 
a negligible factor, comparable to our calibration uncertainty.  

Following our N$_2$H$^+$ treatment, we first solve the NH$_3$ level 
populations assuming no hf splitting and then calculate the emerging
spectrum using sublevel populations proportional to the
statistical weights. To do this, we have modified the Monte Carlo 
code to include the inversion doubling of the rotational
levels in a symmetric top molecule. 
Given the low gas temperature in our cores (9.0-9.5~K), more than 99.99\%
of the para-NH$_3$ population is expected to lie in the 6 levels
resulting from the inversion of the rotational states
(J,K)=(1,1), (2,1), and (2,2), so only those will be considered in our 
models. The energies of these levels
have been calculated with the parameters of \citet{poy75},
and the results have been checked against the values in \citet{pic98}.
The 6 inversion levels are connected by 5 radiative transitions, and their
Einstein coefficients are given by the standard expression for a symmetric
top molecule (see, e.g., \citealt{tow55}) with a dipole moment of 
1.476 debye \citep{coh74}. Collisional coefficients connecting
all levels have been calculated by \citet{dan88} for temperatures 
between 15 and 300 K, and their results show that the de-excitation 
coefficients depend weakly on temperature. Thus, we  
have used for our modeling the 15 K de-excitation
coefficients with para H$_2$,
and calculated the temperature-dependent excitation rates using
the condition of detailed balance for the correct temperature. 
As a final step in our calculation, the emergent intensities have 
been convolved with a Gaussian of $40''$ FWHM, to simulate the 
result of an observation with the 100m telescope.

As mentioned before, a simple LTE analysis of the NH$_3$ data indicates
a constant temperature for all cores. With our Monte Carlo
calculation we can now test this result by fitting the (1,1) line
and checking whether the (2,2) line fits automatically. 
Given that the ratio between the intensity of these lines is very sensitive
to the kinetic temperature \citep{wal83}, any mismatch in 
the (2,2) model will indicate a deviation from the chosen
temperature.

The fit of the C$^{18}$O(1--0), CS(2--1), and 
N$_2$H$^+$(1--0) spectra in the previous sections has already constrained
the turbulent and systematic velocity fields in the models, and this
leaves the molecular abundance as the only free parameter in our NH$_3$ 
calculation. As we did for the other species, we begin the analysis by
running models with constant NH$_3$ abundance and comparing
them with the observed radial profiles. The dashed curves in 
Fig. 6 show that the emission from these models 
is too flat to reproduce the observed central
increase in both the (1,1) and (2,2) transitions, suggesting that 
constant NH$_3$ abundance models are inconsistent with the data
(the L1400K emission can be fit with a constant abundance model,
but our data are too noisy to tell). 

The poor fit of the constant abundance models 
affects both the (1,1) and (2,2) transitions, so
it cannot have resulted from a 
wrong choice in the gas kinetic temperature. It cannot be due 
either to an incorrect estimate of the excitation temperature, as
the (1,1) and (2,2) lines are thermalized to a good approximation 
(gas densities are close to $10^5$ cm$^{-3}$ while the critical density 
is $2 \times 10^3$ cm$^{-3}$). Different tests assuming constant 
excitation temperature or an excitation temperature as given by a two level 
approximation \citep[e.g.,][]{stu85} confirm the Monte Carlo results
and show that fitting the observed radial profiles requires
an enhancement of the central  NH$_3$ abundance.


\begin{figure*}
\figurenum{6}
\epsscale{1.5}
\plotone{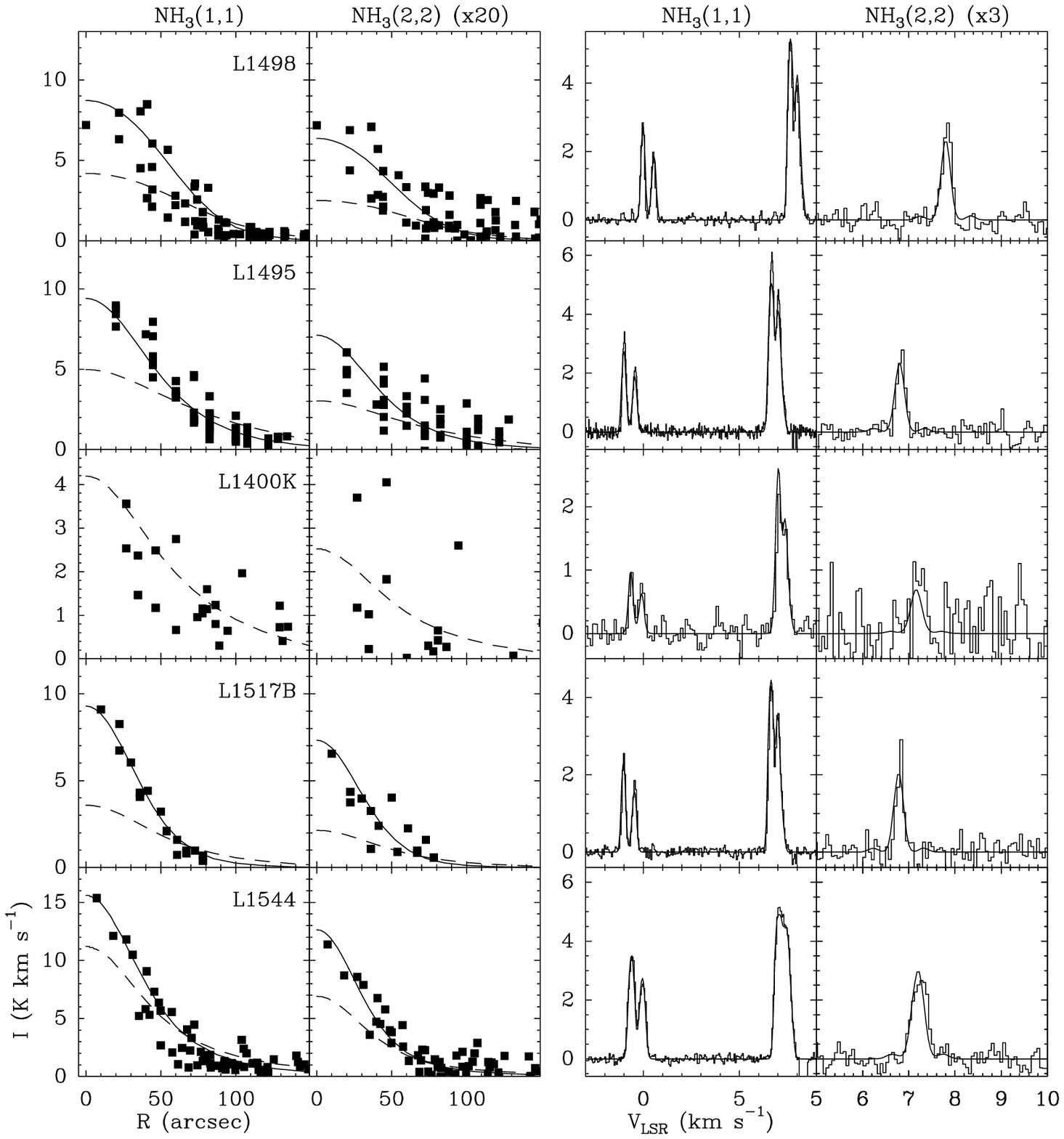}
\caption{
Radial profiles of NH$_3$(1,1) and (2,2) integrated emission (left) and
central emerging NH$_3$(1,1) and (2,2) spectra (right). Observations
are indicated by solid squares in the radial profiles (sum of all hyperfine
components) and by histograms in the spectra (all temperature units are
in the main beam scale). Dashed lines
represent constant abundance models and solid lines represent
models with a central abundance enhancement (see Table 5 for values).
Note how in all
cores but L1400K the models with constant NH$_3$ abundance fail to
fit the central intensity (models have been set to fit outer core
emission), while the models with a central abundance enhancement fit
the data at all radii for both the (1,1) and (2,2) transitions.
For the emerging spectra,
only part of the hyperfine group is presented in order to show in detail the
matching of the linewidth and the hyperfine contrast.
To use the same velocity scale in all spectra, both data and
model for L1400K have been shifted by 4 km s$^{-1}$, and by 1 km s$^{-1}$
for L1517B.
\label{fig6}}
\end{figure*}


To model the NH$_3$ abundance enhancement we have tried different 
analytic expressions, such as power laws and exponential forms. The limited 
resolution of the observations ($40''$) makes it possible to fit the data
with different expressions, as long as the abundance increases by 
a factor of a few toward the center. 
Here we present the results of power law models
$(X(r) = X_0 \; (n(r)/n_0)^\beta),$ shown by solid lines 
in Fig. 6. The values for $X_0$ and $\beta$ are given in Table 5,
which shows that the abundance enhancement $f$ ranges between 3 and 12
(excluding L1400K).

The abundance enhancement models not only fit the (1,1) radial profiles,
but those of (2,2) as well, and this implies a correct choice for
the kinetic temperature of the gas from the LTE analysis. This is not 
surprising given the high degree of thermalization of the NH$_3$ levels, 
which makes the LTE approximation rather accurate.
Although it is still possible
that our $40''$-resolution observations miss gas in the inner cores at
significantly lower temperatures \citep{eva01,zuc01}, the 
sharp central abundance increase, which biases the emission toward the 
innermost gas, 
suggests that this cold region has to be reduced to a radius significantly
smaller than 0.015~pc ($22''$ at 140 pc).

Another satisfactory aspect of the NH$_3$ model is the good fit
to the central spectrum, in particular, to its hf structure.
This can be partially seen in the right panels of Fig. 6, where
the main and the first satellite components of the observed 
(1,1) spectra are compared with the model predictions. Given the 
absence of any free parameter to control the hf ratio of the
model results, it is remarkable that in all cases the model
predicts the correct intensity contrast between hf components, 
and therefore, the correct optical depth. This threefold fit
of the NH$_3$ data (radial profiles of two transitions and hf ratio)
strongly suggests 
that the abundance increase of  NH$_3$ toward the core
centers is an inescapable consequence of the observations.

\section{Discussion}

\subsection{Molecular Depletion and Comparison with Models}

The analysis of the previous sections shows that the cores in our sample
share a similar pattern of chemical differentiation. All systems
present a 
sharp, order of magnitude, central abundance drop in  C$^{18}$O and 
CS, a constant
abundance of N$_2$H$^+$, and a central enhancement of NH$_3$. 
This pattern seems independent of core properties like the gas 
velocity structure (inward/outward), the core size, or its exact shape, 
indicating that it has to arise from a robust and general chemical process.

The most natural explanation for the abundance drops 
of C$^{18}$O and CS is the depletion of these 
molecules onto cold dust grains at high densities. Molecular depletion 
has been widely expected and sought 
\citep[e.g.,][]{mun97}, and our data add to
an increasing number of observations that indicate its final 
detection \citep{kui96,wil98,kra99,alv99,cas99,jes01,ber01}. 
The systematic pattern in our sample cores shows
that molecular depletion is not just limited to a few special cases, but 
that it 
characterizes the majority (or totality) of the starless core population 
at densities larger than a few $10^4$ cm$^{-3}$ (Table 4). 

The coexistence of strong C$^{18}$O and CS depletion with a constant
N$_2$H$^+$ abundance may appear at first surprising, and even more so,
the enhancement of NH$_3$ abundance at the density peak. A likely
explanation for this  behavior comes from the recent work by
\citeauthor{ber97} (\citeyear{ber97}, BL97 hereafter), who have 
presented a chemical model in which differential depletion arises naturally
from the different binding energies of the different species
\citep[see also][]{cha97,aik01,cas01c}. According to BL97,
the N$_2$ molecule binds more weakly to grains than do CO 
and CS, and thus it is more easily desorbed. This difference in
binding energies gives 
rise to a pattern of depletion in which N-rich molecular species
survive in the gas phase at higher densities than do CO and CS, which  
disappear at densities of about $10^4$ cm$^{-3}$. To further explore 
the BL97 model predictions, we have used our radiative
transfer calculation to estimate emerging
intensities for the cores in our sample.

        BL97 have studied the chemical evolution of a gas parcel as its
density increases with time following two different models, 
one based on theoretical work by  \citeauthor{bas94} (\citeyear{bas94}, 
BM94 hereafter), 
and the other based on a phenomenological (accretion) fit to the contraction
history of L1498 \citep{kui96}. The main difference between these two 
models is that the BM94 contraction is rather slow (as it is driven by 
ambipolar diffusion), while in the accretion fit model the density 
initially increases much faster. For both models, BL97 studied the effect
of two types of grain mantles (CO and H$_2$O), but 
a look at their abundance predictions shows that 
significant CO depletion only occurs in models having dust grains with 
(tightly bound) H$_2$O mantles. Thus, in the following discussion we 
concentrate on models with these types of grains.

To compare the BL97 predictions with our observations of real cores,
we create a model core with the density distribution of one of
our observed cores and an abundance distribution that matches point
by point the abundances predicted by BL97.
Unfortunately, the BL97 models never exceed densities larger than 
3-5 $\times 10^4$~cm$^{-3}$, 
so no abundance predictions are given for the densities typical of 
our core centers (1-2 $\times 10^5$~cm$^{-3}$, Table 2). This is immaterial
for CS and CO, since they are so depleted at these densities that 
the exact value of the abundance does not affect the radiative
transfer results. However, it
is critical for the N$_2$H$^+$ and NH$_3$ analysis, as some BL97 models
show the beginning
of an N$_2$H$^+$ and NH$_3$ abundance decrease at the highest densities (most
likely due to depletion), and this makes any extrapolation to higher
densities especially unreliable. For those molecules 
we have explored two possibilities:
that the abundance stays constant at the highest densities
(unlikely according to BL97 Figs. 2 and 3), and that the abundance drops
by an order of magnitude for densities not studied by BL97 (a conservative 
depletion model).


\begin{figure*}
\figurenum{7}
\epsscale{1.8}
\plotone{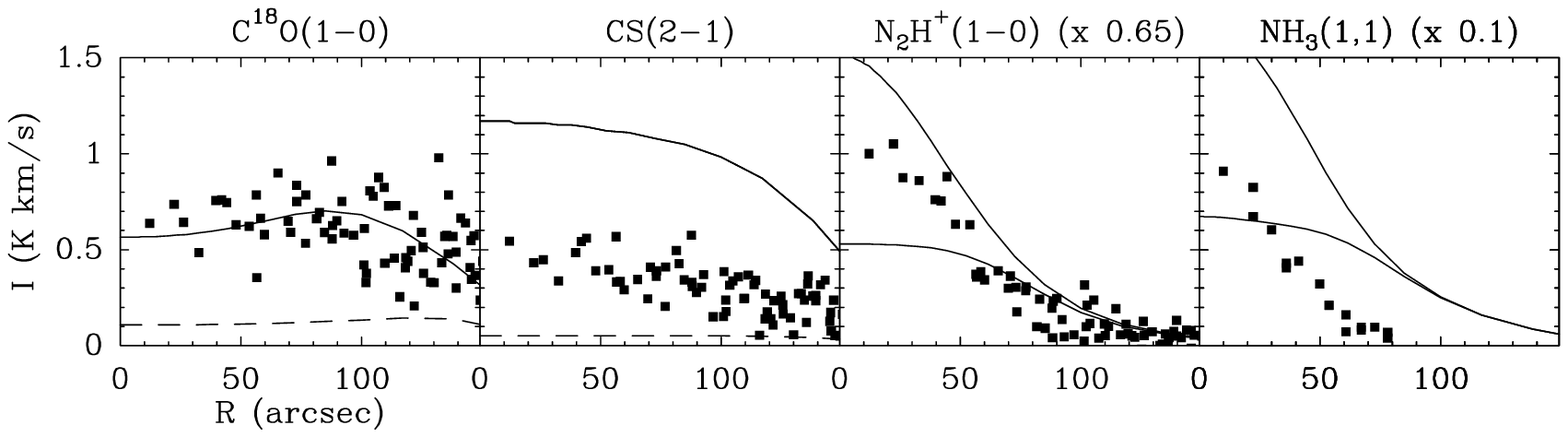}
\caption{
Comparison between L1517B observations (solid squares) and expected emission
from \citet{ber97} models (lines). In the C$^{18}$O and CS
panels, the dashed lines correspond to the BM94 model (their Model 1), and the
solid lines indicate the accretion model (Model 2). In the
N$_2$H$^+$ and NH$_3$ panels, only results for the accretion model
are shown. The top lines in these panels represent no depletion,
and the bottom lines represent a factor of 10 depletion for densities
higher than those studied by \citet{ber97} (see text).
All models assume dust grains with H$_2$O mantles, and to
simplify the display, both model and data in the two rightmost
panels have been scaled down by 0.65 (N$_2$H$^+$) and 0.1 (NH$_3$).
\label{fig7}}
\end{figure*}


Fig. 7 compares the BL97 model predictions for the L1517B core 
with our observations of C$^{18}$O, CS, N$_2$H$^+$, and NH$_3$ (a 
similar comparison with L1498 gives the same result).
As can be seen, the BM94 contraction history (dashed lines) 
predicts C$^{18}$O and CS 
intensities 5 times weaker than observed, while the accretion model 
predicts the 
correct C$^{18}$O intensity and overestimates the CS emission by a factor of 
3. The weaker C$^{18}$O and CS emission in the BM94 model result from its
slow core contraction, which makes the outer core material 
rather old (several Myr) and, 
therefore, heavily depleted of these species. The 
accretion model, on the other hand, predicts a younger outer core 
($<1$ Myr) with much less depletion. The higher CS emission predicted
by the accretion model arises from an overestimate 
of the undepleted CS abundance by at least an order of magnitude.

Both the BM94 and accretion models predict similar N$_2$H$^+$ and NH$_3$
abundances, so Fig. 7 only presents the results for the accretion model.
The higher curves in the figure are intensity predictions for the case 
of no depletion even at the highest densities, while the lower curves
assume a moderate depletion of a factor of 10 for densities higher 
than those studied by BL97. As can be seen, the models with no depletion 
agree much better with observations, although these 
models are less likely according to the BL97 plots.
The undepleted N$_2$H$^+$ model is particularly good, and it
predicts the correct intensity within 25\% (again, a possible decrease 
in trapping by
hfs was neglected). The  NH$_3$ prediction, on the other hand,
not only overestimates the intensity, but it does not reproduce the
NH$_3$ abundance increase observed toward the core centers
(an ortho-para ratio of 2 was assumed). This is a 
general problem of the BL97 models, which predict slight
NH$_3$ abundance increases at late times, but they are too small
to fit the NH$_3$ enhancement inferred from our 
observations. A possible solution is suggested by the 
recent work of \citet{nej99}, whose models show
a sharp enhancement of the NH$_3$ abundance in the core
inner 0.1 pc while the N$_2$H$^+$ fraction remains constant.
More detailed modeling of the chemical reactions triggered by the 
disappearance from the gas phase of CO and other species is clearly
necessary.

If we take the BL97 models at face value, the previous discussion
implies that cores condense rapidly and have dust
grains coated with H$_2$O ice. These conclusions, however, depend critically
on the assumed molecular binding energies, and they 
may change as more accurate molecular parameters are derived.
A lower CO binding energy, for
example, will slow down the rate of CO depletion, and this can
favor the larger evolution time scales predicted by 
ambipolar diffusion. Grain properties, in addition, may evolve 
as molecules deplete. An initially H$_2$O-coated grain may
evolve (through CO depletion) into a CO-coated grain, and its
depletion ability may obviously change, or as
\citet{cas01c} have proposed, 
the presence of atomic oxygen in the gas phase may maintain
polar mantles in the densest parts of cores, given that a fraction of
O will deplete and quickly be transformed in H$_2$O. These and other 
uncertainties
show that it is premature to infer core evolution properties
from the observed pattern of abundances. Still, the reasonable success
of the BL97 model (and those by \citealt{aik01} and \citealt{cas01c})
shows that most, or all,
the chemical differentiation we have observed in starless cores very likely
arises from depletion and reactions triggered by depletion.

\subsection{A Solution to the NH$_3$/CS Size Discrepancy}

The systematic chemical differentiation of cores we have found, and
particularly the CS abundance drop, suggests a natural explanation 
for the decade-old discrepancy between CS and NH$_3$ observations 
\citep[e.g.,][]{ful87,zho89,pas91,mor97}. This discrepancy arises because
CS rotational transitions have larger critical densities than NH$_3$ 
inversion transitions. Thus, when mapping a centrally concentrated core, 
one would expect the CS emission to appear more peaked
and concentrated
than the NH$_3$ emission. In practice, however, CS maps are systematically 
more extended than NH$_3$ maps, with ratios between CS and NH$_3$
diameters of 
$1.5\pm0.3$ \citep{zho89} or $\approx 2$ \citep{mye91}.

Explanations for the larger extension of the CS emission have 
postulated the scattering of photons by the core outer
gas \citep{ful87}, CS abundance enhancement by
shocks in the outer core layers \citep{zho89}, or unresolved clumps 
\citep{tay96}. However, it has not been proved that these mechanisms 
are truly at work, and the
NH$_3$/CS discrepancy has remained unsolved for more than a decade.
In this section we explore quantitatively how molecular 
differentiation helps to explain the discrepancy, and for this purpose,
we use our Monte Carlo
model results as if they represented observed data and analyze them
using standard procedures.

Our models, by construction, fit simultaneously the emission size
of C$^{18}$O(1--0), CS(2--1), N$_2$H$^+$(1--0), NH$_3$(1,1) 
and NH$_3$(2,2) for each of the 5 cores in our sample. 
This means that these models should naturally give rise to 
any difference in emission size between tracers and, therefore, they
should reproduce the NH$_3$/CS discrepancy. 
The advantage of reproducing the discrepancy with the models is that
the internal model parameters are well known, so we can use the models
as laboratories to explore how the
discrepancy arises from the core physical properties. Taking
advantage of our multi-line observations, we
study the behavior of C$^{18}$O and N$_2$H$^+$, in addition to CS
and NH$_3$.

\begin{figure*}
\figurenum{8}
\epsscale{1.8}
\plotone{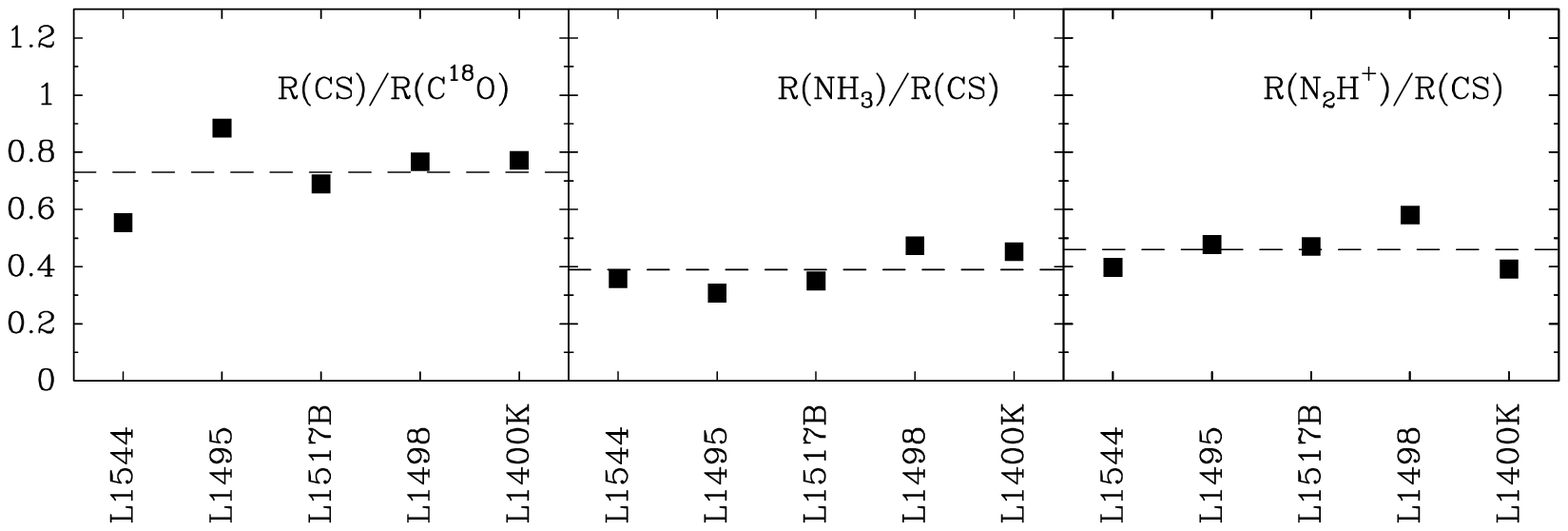}
\caption{
Ratio of map sizes (half-maximum radius) for different
molecules as measured from
the best-fit Monte Carlo models represented in Figs. 3, 5, and 6.
The dashed lines represent the mean value for each molecular
pair (see text). Note how the NH$_3$ sizes are systematically
smaller than the CS sizes by a factor larger than 2. This
systematic size discrepancy, already known but unexplained,
is a consequence of the central abundance drop of CS,
which severely truncates the emission from this molecule
and therefore increases the half-maximum radius. See text for
a complete discussion.
\label{fig8}}
\end{figure*}

To estimate the sizes of the emission in different molecules, we 
treat our model results as if they were real data taken
with a telescope (FCRAO or 100m Effelsberg), and follow the standard 
procedure 
of measuring sizes using the half maximum radius \citep{zho89,mye91}.
For NH$_3$, N$_2$H$^+$, and CS this process is easily
done directly from the model radial profiles, but for C$^{18}$O,
some radial profiles present a slight central dip, so the peak intensity
does not correspond to the value at the core center. For compatibility 
with the
other molecules, we define as radius the distance at which the
C$^{18}$O intensity drops to half the value at the core center.
Had we taken the C$^{18}$O peak value (off center), the radius would have
been larger, but only by a very small amount (e.g., 5\% 
in L1517B).

Figure 8 shows the CS/C$^{18}$O, NH$_3$/CS, and N$_2$H$^+$/CS
size ratios measured from our Monte Carlo models following the above
procedure. As the figure shows, the CS radii are consistently
smaller than the C$^{18}$O radii by about 30\%,
and the NH$_3$ and N$_2$H$^+$ radii
are consistently smaller than the CS radii by a factor of 2 or more.
Quantitatively, we find that
$\langle R(CS)/R(C^{18}O)\rangle = 0.73 \pm 0.12$,
$\langle R(NH_3)/R(CS)\rangle = 0.39 \pm 0.07$,
and
$\langle R(N_2H^+)/R(CS)\rangle = 0.46 \pm 0.08$.
(A direct estimate of these ratios from the sizes of
the half maximum contours in the maps of Fig. 1 yields 0.77, 0.39,
and 0.54 for $R(CS)/R(C^{18}O)$, $R(NH_3)/R(CS)$, and $R(N_2H^+)/R(CS)$,
respectively, in excellent agreement with the values estimated from the 
Monte Carlo models.)

The value of $\langle R(CS)/R(C^{18}O)\rangle$ in our Monte Carlo models
is in excellent agreement with the
$\approx 0.8$ derived by \citet{mye91} from their sample of
cores, and the $\langle R(NH_3)/R(CS)\rangle$ value also agrees 
with the $\approx 0.5$ measured by \citet{mye91}. 
This good match between the relative sizes of        
the Monte Carlo models and those of the
general population of cores suggests  that the models not only reproduce
the details of our (small) core sample, but that they capture the basic
properties of the general core population (both starless and with stars). 
In particular, 
the models automatically reproduce the CS/NH$_3$ size discrepancy.

To understand how the CS/NH$_3$ size discrepancy arises in the models 
(Fig. 8 central panel),
we isolate the factors that cause the NH$_3$ maps to be 
smaller and the CS maps to be larger, and look for
the dominant one. The central NH$_3$ abundance enhancement
obviously decreases the size of the NH$_3$ maps,
and we quantify this effect by comparing the radii of the best fit
with that of the constant abundance models in Fig. 6. In this way, we
find that best-fit models (i.e., abundance enhanced) 
are about 30\% smaller than constant-abundance models, which is 
the right trend  but is not 
enough to explain the CS/NH$_3$ size discrepancy. 
The small effect of the abundance enhancement
is consistent with the fact that
there is a similar size discrepancy between CS and
the constant abundance N$_2$H$^+$ (see right panel in Fig. 8).

Two factors that increase the size of the CS maps are 
optical depth and chemical differentiation. We study the effect of optical
depth by decreasing the CS abundance with a global factor of 22 (to simulate 
a thin C$^{34}$S observation), and find that this only decreases
the average half-maximum radius by a factor of 20\%. This result is 
consistent with the C$^{34}$S observations of L1544 (Fig. 4),
which show that the rare-isotopomer map is only 25\% smaller than the
CS map, despite the fact that the main isotope is extremely thick and 
self absorbed. We, therefore, conclude that optical depth is
only a minor factor in the size discrepancy.

Finally, we explore the effect of the central CS abundance drop by
comparing the sizes of our best CS models with the sizes predicted by
constant CS abundance models which fit the observed central intensity.
These constant-abundance models are a full factor of 2 smaller
than the models with abundance drop, so they have sizes more comparable to 
those of NH$_3$ and N$_2$H$^+$. This means that the major contributor 
to the CS/NH$_3$ model discrepancy
is the central drop in CS abundance, which decreases the peak CS intensity
by a factor of several (see Fig. 4 bottom panel), and this in turn 
increases the radius of the half-maximum contour by about a factor of 2.

\subsection{The NH$_3$/CS Linewidth Discrepancy and 
the Linewidth-Size Relation in Starless Cores}

\begin{figure*}
\figurenum{9}
\epsscale{1.0}
\plotone{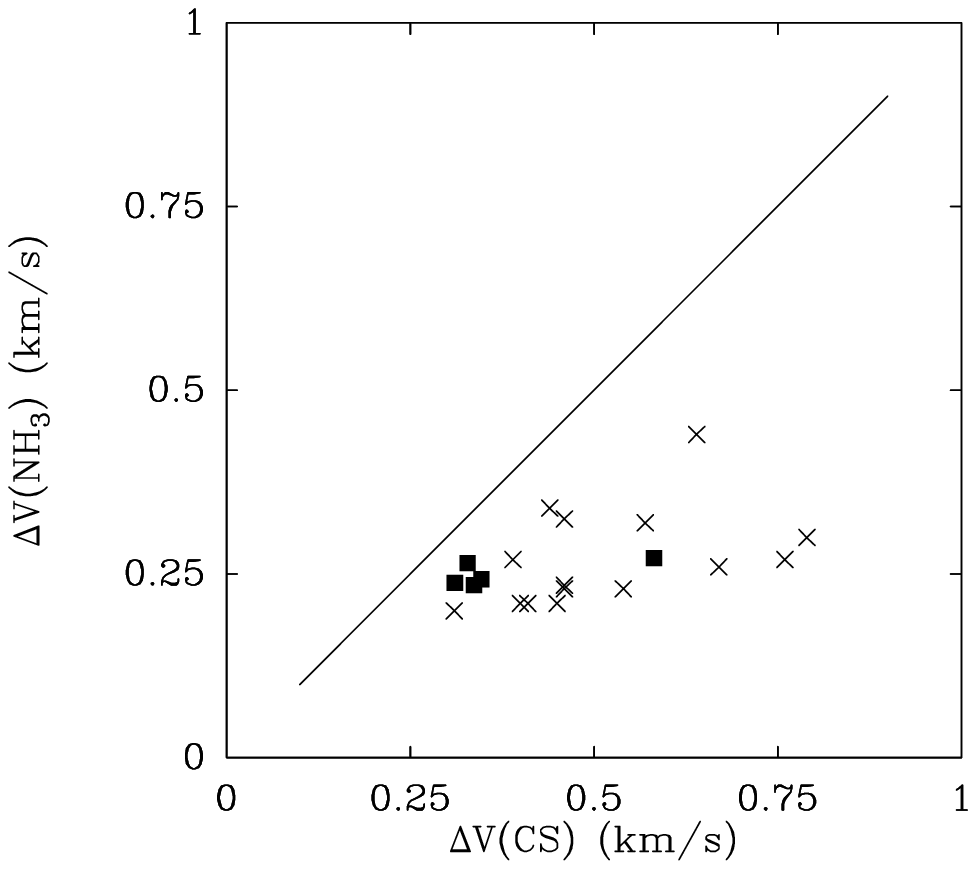}
\caption{
Comparison of CS and NH$_3$ linewidths from dense cores. Crosses
are data from \citet{zho89} (starless cores only), and filled
squares are results from our Monte Carlo models.
Both real data and models lie below the line of equal CS and NH$_3$
linewidth, implying that CS lines are systematically broader than
NH$_3$ lines. Although this linewidth discrepancy has been traditionally
interpreted as resulting from the systematic increase of gas turbulence
with core radius, none of our models has this property.
Optical depth broadening and self absorption are the two causes of
the broader CS lines in the Monte Carlo models (see text).
\label{fig9}}
\end{figure*}

A problem commonly associated with the NH$_3$/CS size discrepancy is 
the systematic difference in linewidths between these two dense gas
tracers. In their comparison between NH$_3$ and CS observations of 
dense cores, \citet{zho89} found that 
$\Delta V(CS)/\Delta V(NH_3) = 2.0\pm 0.6$, which is too large to be
explained by optical-depth broadening of the CS lines. Given the 
additional larger size of the CS emission, the traditional  
interpretetaion of this effect has been in terms of a systematic increase
in the gas turbulence with radius, following the so-called
linewidth-size relation \citep[e.g.,][]{ful92}.

Given the above explanation, 
it appears surprising that all our cores (but L1544) 
are well fit by models with constant turbulent velocity, and that
for L1544, our fit requires a {\em decrease} of linewidth
with radius, not an increase. This result is significant 
because the line modeling is very sensitive to any linewidth variation
along the line of sight
through the simultaneous fit of the CS(2--1) self absorption
(which arises from gas at large radius) and the fit of the 
N$_2$H$^+$ and NH$_3$ lines (which represent gas from
the innermost core). The success of the constant-linewidth models
suggests that, in most cores, 
there are no systematic variations in the turbulent-velocity
field from the core center up to radii of at least 0.1 pc (core edge).

To study whether our modeling is consistent with the NH$_3$/CS linewidth 
discrepancy observed in other cores, we again use our model data as if they 
were real observations, and measure CS and NH$_3$ linewidths
from the predicted spectra. To do this, 
we first convert the model spectra into CLASS format and then
use that program to fit the lines. Following standard procedure, we 
fit the CS(2--1) lines with simple Gaussians, and the NH$_3$(1,1) multiplets
with the standard hyperfine analysis. We apply this procedure even to
the obviously self-absorbed L1544 CS spectrum, and in Fig. 9 plot our model
data (black squares) together with the data from \citet{zho89} for
starless cores (crosses). No attempt
has been made to correct our model data from instrumental broadening,
as the resolution of our model ensures that this effect is less
than 5\%.

As Fig. 9 shows, our model data lie in the same region where we find 
the data from observations of other cores, i.e. consistently
below the line of equal NH$_3$ and CS linewidth. This shows that our
models naturally display the NH$_3$/CS linewidth discrepancy, the
way they did for the size discrepancy. The model data, in addition,
predict $\Delta V(CS)/\Delta V(NH_3) = 1.5\pm 0.4$, which is consistent
with the ratio derived by 
\citet{zho89}. The clustering of most of our points in the 
region of smallest linewidth is probably due to the fact that we 
selected the most quiescent cores possible (Section 2), and in fact,
all cores but L1544 were modeled
using the same turbulent speed (L1544 is the outlying point with large 
$\Delta V(CS)$). From Fig. 9 and the measured linewidth ratio, we conclude
that a significant 
NH$_3$/CS linewidth discrepancy can be produced even in the 
presence of a constant turbulence velocity field, and that it does not
require the presence of a systematic increase of linewidth with size.

To understand how this NH$_3$/CS linewidth discrepancy arises in
constant turbulence gas, we have analyzed how lines are generated in our 
models and have found two main factors that contribute
to the apparent increase of the CS linewidth over that of NH$_3$.
First, there is a bias in the way the two linewidths are 
measured: in the standard procedure,
CS linewidths are derived by fitting a single Gaussian while 
NH$_3$ linewidths are estimated  with a hyperfine
fit. This method has the somewhat perverse effect of correcting
he usually not-very-thick NH$_3$ line for opacity broadening,
while not correcting at all the always-thick CS spectra. As 
\citet{zho89} mention, this effect contributes to, but cannot be the 
only factor in, the NH$_3$/CS linewidth discrepancy.

The additional factor that enhances the CS linewidth is the 
presence of self absorption. Self absorption partially truncates the CS 
line, decreasing its peak intensity and systematically increasing 
the size of its half-maximum width. This effect is
similar to the way the depletion of CS at the core center 
truncates the map of emission and increases its half-maximum radius.
Together with the linewidth-measurement bias, it is responsible for 
the consistently larger CS linewidths in our models (systematic velocity 
fields have only a minor effect). 

The effect of self absorption
in the final linewidth is very sensitive to the velocity displacement
between the emitting and absorbing gas. A large-enough displacement (like 
in L1498 and L1517B, see Fig. 3) produces a single-peaked line slightly
broader than the NH$_3$ line, while no displacement (like in L1544) produces
a double-peaked line which yields an extremely large linewidth when fitted
with a Gaussian. This effect shifts the L1544 point very far to the right in
Fig. 9, and it suggests that the points with large $\Delta V(CS)$ 
from \citet{zho89} represent spectra with deep central self-absorptions,
although further work is needed to clarify this issue.
(Note that the strongly double-peaked CS line
from L1544 would have appeared rather diluted in the 0.09 km s$^{-1}$
channels of \citealt{zho89}, and that these authors recognize that
some of their lines are truly self-absorbed.)

Our conclusion that the gas turbulence does not increase with radius 
in the inner part of our cores, based on the analysis of 
line-of-sight motions, agrees with recent studies of turbulent
motions in the plane of the sky for other cores. Using NH$_3$ as a 
single tracer, \citet{bar98} find negligible
linewidth variations up to distances of 0.1 pc, which is inconsistent 
with the factor-of-3 increase expected from
the standard linewidth-size relation between 0.01 and 0.1 pc.
This suggests that the common interpretation of a turbulence
increase in the inner 0.1 pc of some cores, 
based on the combination of NH$_3$ and CS data,
may arise from the combination of optical
depth effects (which increase the CS linewidth) and a central
abundance drop (which increases the CS half-maximum radius).

What happens at radii larger than 0.1 pc is still unclear.
Large-scale studies show that there is a true
linewidth-size relation when considering whole clouds 
\citep[e.g.,][]{lar81}, and \citet{goo98} have suggested that turbulence
starts to increase at radii larger than 0.1 pc. Unfortunately
their NH$_3$ data are not sensitive enough to follow this trend very far.
An alternative view suggested by some of our CS and C$^{18}$O data is that the
large-scale broadening may result from the overlap of different cloud 
fragments moving at different velocities. Further work on large-scale 
modeling of molecular lines should be carried out to clarify this 
crucial topic.

\subsection{Further Consequences of Chemical Differentiation}

The chemical differentiation of starless cores has additional
implications for studies of low-mass star formation, and in this 
section we briefly discuss some aspects of this problem.
CS, for example, is one of the molecules of choice for density 
determinations (e.g., 
\citealt{eva99}), thus, its vanishing from the gas phase at densities of 
a few 10$^4$ cm$^{-3}$ implies that any attempt to 
measure with this molecule
the density profile of a concentrated core seems doomed to miss the inner
core region and, therefore, to underestimate the central density. 
This same problem may affect
other classical density tracers like HCO$^+$, HCN, and H$_2$CO
(see BL97 for model predictions), and a systematic analysis is in progress
for these and other molecules (Tafalla et al., in preparation).
The constant abundance of N$_2$H$^+$ suggests that this molecule 
may be a robust density tracer (see \citealt{cas99}), but 
its complicated hyperfine structure and a lack of accurate collisional 
rate coefficients make its radiative transfer much more difficult to
solve. Although it may be argued that dust continuum observations 
may suffice to derive core density profiles (our approach in section 4.2), 
confirmation of these estimates with molecular observations seems 
necessary, given the present uncertainties in the dust emission 
properties. 

An additional issue in the study of starless cores potentially affected
by chemical differentiation is the search for infall motions. Infall 
searches are usually done combining
observations of optically thick and thin tracers, often
from different molecular species (e.g., \citealt{lee99}).
A common choice consists of CS (thick) and N$_2$H$^+$ (thin), two species 
which we have shown to reside in
different parts of the core. This disjoint distribution of
tracers complicates the interpretation 
of the spectral signatures and can only be avoided
using isotopomer pairs of different optical depth. The sharp
central abundance drop of CS, in addition, implies that infall searches 
using this molecule are only sensitive to the motions of the core's 
outer layers. Thus, to probe motions close to the core center, 
a molecular species
resistant to differentiation/depletion, like N$_2$H$^+$, is  
needed. The lower optical depth of the N$_2$H$^+$ lines, however, 
complicates detecting inward motions with the standard self-absorption
analysis (but see \citealt{wil99} for a claim of N$_2$H$^+$ self-absorption
in L1544). The search for high velocity wings in the lines of
this tracer appears to be a complementary and promising technique 
\citep{bou01}.

Chemical differentiation does not only represent 
an added difficulty in the study of starless cores.  
The time dependence of the underlying chemical processes
suggests that detailed chemical modeling should allow us
to reconstruct core contraction histories
from observed depletion patterns. We cautioned
in section 6.1 that the fast contraction scenario derived
from the BL97 model is still tentative, as it depends on the
exact value of the molecular binding energies, and that an 
accurate determination of these critical parameters is still
needed to definitively test the longer time scales
predicted by ambipolar diffusion models. Before such
quantitative time estimates are possible, we can compare the 
abundance patterns in different cores in an attempt to construct a 
qualitative time sequence of core evolution.
Most cores in our sample have similar central densities
($\sim 10^5$ cm$^{-3}$) and similar depletion patterns
(Tables 4 and 5), so it seems likely that they are at a 
similar evolutionaly stage (which may be just an effect
of our selection criteria). Interestingly enough,
our sample contains cores with both inward and
outward motions (L1498 and L1517B, respectively), suggesting
that the presence of these large-scale velocity patterns
has little to do with the core evolutionary state.

The only clue to an evolutionary pattern in our sample comes from 
the most extreme object, L1544. This core has density higher than the
others by an order of magnitude 
($\sim 10^6$ cm$^{-3}$), and its abundance pattern slightly 
differs in the sense that the abundance variations of
CS, C$^{18}$O, and NH$_3$ occur at higher densities 
than in the other cores (Tables 4 and 5).
This could mean that a very rapid contraction 
has kept frozen a chemical pattern characteristic of lower densities or that
the core has contracted faster than the others, which would not allow 
enough time for the moderately dense gas to be chemically processed.

Clearly additional work is needed before going further with the
above speculations. The chemical composition of cores
with different densities and conditions should be measured in 
detail. Further modeling of the chemical evolution of cores as
they contract is also needed. Finally, it is important
to identify key molecules that allow us to distiguish
between different chemical scenarios, and to characterize the
chemical behavior of standard high-density tracers.
The potential benefits of understanding 
the chemical evolution of cores make exploring these and 
related questions an urgent enterprise.

\section{Conclusions}

We have mapped 5 starless cores (L1498, L1495, L1400K, L1517B, and L1544) 
in C$^{18}$O(1--0), CS(2--1),  
N$_2$H$^+$(1--0), NH$_3$(1,1) and (2,2), C$^{17}$O(1--0) (3 cores),
C$^{34}$S(2--1) (1 core),
and the 1.2 mm continuum (4 cores), and complemented these data with
a published 1.2mm continuum map of L1544 \citep{war99}.
For each core, we have self-consistently determined
the radial profile of density (from the 1.2mm continuum),
temperature (from NH$_3$), and molecular abundance (using a 
radiative transfer Monte Carlo model). To do this, we 
have fit simultaneously
the central spectrum for each line and the radial profile of integrated 
intensity. As a result of this work, we conclude the following:

1. Each core is well fit with a radial density profile of
the form $n(r) = n_0/(1+(r/r_0)^\alpha)$ where $n_0$ ranges from
about $10^5$ to $10^6$ cm$^{-3}$, $r_0$ is of the order of 3000-10000 AU, 
and $\alpha$ ranges from 2 to 4. This type of profile naturally fits the 
central flattening and the large-$r$ power-law behavior found by previous 
authors.

2. All cores show evidence for a strong central drop in the abundance of 
C$^{18}$O and  CS. This drop can be fitted with a negative exponential 
dependence on
density ($\exp (-n(r)/n_d)$), with an e-folding $n_d$ parameter 
of 2-6 $\times 10^4$ cm$^{-3}$. C$^{17}$O(1--0)
and C$^{34}$S(2--1) observations of several cores confirm these results
and show that our radiative transfer models can naturally fit this
rare isotopomer emission without extra free parameters. 
Although limited by the resolution of our data 
($\sim 50''$), we estimate that the C$^{18}$O and  CS central abundance 
drops are at
least of one or two orders of magnitude in each of the cores.

3. In contrast to C$^{18}$O and CS, the N$_2$H$^+$ abundance seems to 
be constant inside each core.
The combination of narrow N$_2$H$^+$(1--0) central
spectra and the presence of narrow CS(2--1) self absorptions
indicate that the turbulent linewidth in each core is constant with radius.

4. The NH$_3$(1,1) and (2,2) lines are well fit with constant 
temperature models of 9.5 K (8.75 K in L1544) 
in which the NH$_3$ abundance increases 
toward the core center. The fact that the temperature is constant at 
densities for which gas and dust may be thermally coupled
supports the assumption of constant dust temperature used in 
our 1.2mm continuum analysis.

5. The combination of a central C$^{18}$O and CS abundance drop,
a constant N$_2$H$^+$ abundance, and a central NH$_3$ abundance
increase seems to be a systematic characteristic of starless cores. 
This suggests that star-forming material is very chemically
inhomogeneous before starting to collapse, and that searches
for infall in starless cores should take this fact into account.

6. The pattern of chemical differentiation, especially for C$^{18}$O, 
CS, and  N$_2$H$^+$ is in qualitative agreement with depletion
models. Quantitative agreement between 
the data and these models can be achieved for certain choices of
their parameters, which suggest the presence of polar ices on grains and
a history of core contraction more rapid than predicted by subcritical
ambipolar-diffusion models.
These models, however, seem to overestimate N$_2$H$^+$ depletion at
high densities and to miss the observed central NH$_3$ abundance increase.

7. The presence of chemical differentiation in cores 
automatically explains the systematic difference between
map sizes of different tracers, a problem which has remained unexplained
for more than a decade. Detailed Monte Carlo radiative transfer
models show that CS maps, because of the central abundance decrease
and radiative transfer effects, are expected to appear at least twice as 
large as NH$_3$ and N$_2$H$^+$ maps. These latter molecules are therefore 
more faithful tracers of the dense core material.

8. Comparing the CS and NH$_3$ linewidths from our Monte Carlo 
radiative transfer calculations, we find a systematic NH$_3$/CS
linewidth discrepancy similar to that found by other authors 
\citep[e.g.,][]{zho89}. These linewidth discrepancies
have been traditionally interpreted as resulting from a systematic
increase in the turbulent linewidth with radius, but our models show 
that it can arise from a combination of optical depth and self
absorption in the CS line. This suggests that the linewidth-size relation
does not apply in the inner 0.1 pc of cores.

\acknowledgements We thank the staffs of the FCRAO, 30m, and 100m telescopes
for their support. Mark Heyer of FCRAO was always just a phone
call away when any problem arose, and his help is greatly appreciated.
We also thank Peter Schilke for help with the calibration of the 100m data,
Fr\'ed\'erique Motte for providing us with the 1.2mm image of L1544, 
Carl Gottlieb for communicating his new line frequency measurements prior to
publication, and Dale Fuchs and an anonymous referee for their careful reading
of the manuscript and their suggestions to improve the presentation. 
This research has made use of NASA's Astrophysics Data System
Bibliographic Services and the SIMBAD database, operated at CDS,
Strasbourg, France. MT was partially supported by grant AYA2000-0927 from
the Spanish DGES. PC and CMW acknowledge travel support
from ASI Grant 98-116, as well as from the MURST project ``Dust and
Molecules in Astrophysical Environments.''

\clearpage 

\begin{deluxetable}{lcc}
\tablecolumns{3}
\tablecaption{Sample Cores. \label{tbl-1}}
\tablewidth{0pt}
\tablehead{
\colhead{Core} & \colhead{$\alpha(1950)$}  & \colhead{$\delta(1950)$} \\
\colhead{} & \colhead{$ ^{\rm h~~m~~s} $} &
\colhead{$^\circ\phantom{45}' \phantom{30}''$}
}
\startdata
L1498 & 04~07~50.0 & 25~02~13 \\
L1495 & 04~11~02.7 & 28~00~43 \\
L1400K & 04~26~51.0 & 54~45~27 \\
L1517B & 04~52~07.2 & 30~33~18 \\
L1544 & 05~01~14.0 & 25~07~00 \\
\enddata
\end{deluxetable}

\begin{deluxetable}{lccccccc}
\tablecolumns{7}
\tablecaption{Density fits from continuum data\tablenotemark{(a)} \label{tbl-2}}\tablewidth{0pt}
\tablehead{
\colhead{Core} & \colhead{Center\tablenotemark{(b)} }  &
\colhead{$T_d$\tablenotemark{(c)} } & \colhead{$b/a$\tablenotemark{(d)}}   &
\colhead{PA\tablenotemark{(e)}} & \colhead{$n_0/10^5$}  & \colhead{$r_0$} & \colhead{$\alpha$}\\
\colhead{} & \colhead{$('','')$} & \colhead{(K)} & \colhead{} &
\colhead{$(^\circ)$}  & \colhead{(cm$^{-3})$} & \colhead{$('')$}
}
\startdata
L1498 & (0,0) & 9.5 & 0.6 & -40 & 1.0 & 75 & 4\\
L1495 & (20,40) & 9.5 & 1.0 & 0  & 1.1 & 45 & 2\\
L1400K & (-20,-40) & 9.5 & 0.3 & 45 & 1.2 & 30 & 2\\
L1517B & (-10,-20) & 9.5 & 1.0 & 0 & 2.2  & 35 & 2.5\\
L1544 & (-13,-21) & 8.75 & 0.6 & -45 & 14 & 20 & 2.5\\
\enddata

\tablenotetext{(a)} {$n(r) = n_0/ [1+(r/r_0)^\alpha] $}
\tablenotetext{(b)} {Offset with respect to coordinates in Table 1.}
\tablenotetext{(c)} {Dust temperature, assumed equal to gas temperature
(see text for details).}
\tablenotetext{(d)} {Aspect ratio for elliptical average.}
\tablenotetext{(e)} {Position angle for elliptical average.}
\end{deluxetable}

\begin{deluxetable}{lccc}
\tablecolumns{4}
\tablecaption{Core velocity parameters. \label{tbl-3}}
\tablewidth{0pt}
\tablehead{
\colhead{Core} & \colhead{V$_{LSR}$}  & \colhead{$\Delta V$}
& \colhead{$dV/dr$} \\
\colhead{} & \colhead{(km s$^{-1}$)} & \colhead{(km s$^{-1}$)} &
\colhead{(km s$^{-1}$ pc$^{-1}$)}
}
\startdata
L1498 & 7.80 & 0.17\tablenotemark{(a)} & -0.47 \\
L1495 & 6.80 & 0.17\tablenotemark{(a)} & 0.93 \\
L1400K & 3.16 &0.17\tablenotemark{(a)} & 1.55 \\
L1517B & 5.78 & 0.17\tablenotemark{(a)} & 0.78 \\
L1544 & 7.20 & 0.20\tablenotemark{(b)} & 0.00 \\
\enddata
\tablenotetext{(a)}{Spatially constant.}
\tablenotetext{(b)}{Variable, see section 5.1.}
\end{deluxetable}

\begin{deluxetable}{lcccccc}
\tablecolumns{7}
\tablecaption{Best-Fit Model Parameters for C$^{18}$O and CS.
\tablenotemark{(a)} \label{tbl-4}}
\tablewidth{0pt}
\tablehead{
\colhead{} & \colhead{}  & \colhead{C$^{18}$O } & \colhead{}   &
\colhead{} & \colhead{CS}  & \colhead{} \\
\colhead{CORE} & \colhead{$X_0$} & \colhead{$n_d$ (cm$^{-3}$)} &
\colhead{$f$\tablenotemark{(b)}}
& \colhead{$X_0$} & \colhead{$n_d$ (cm$^{-3}$)}  &
\colhead{$f$\tablenotemark{(b)}}
}
\startdata
L1498 & $1.7 \; 10^{-7}$ & $2.0 \; 10^4$ & $2 \; 10^{-2}$ & $3 \; 10^{-9}$  &
$3.0 \; 10^4$  & $8 \; 10^{-2}$ \\
L1495 & $1.7 \; 10^{-7}$ & $2.2 \; 10^4$  & $4 \; 10^{-2}$  & $5 \; 10^{-9}$ &
$1.5 \; 10^4$ & $8 \; 10^{-3}$ \\
L1400K & $1.7 \; 10^{-7}$ & $1.4 \; 10^4$ & $5 \; 10^{-3}$  & $9 \; 10^{-9}$ &
$1.5 \; 10^4$ & $7 \; 10^{-3}$ \\
L1517B & $1.7 \; 10^{-7}$ & $2.1 \; 10^4$ & $4 \; 10^{-4}$  & $3 \; 10^{-9}$ &
$4.0 \; 10^4$ & $2 \; 10^{-2}$ \\
L1544 & $1.7 \; 10^{-7}$ & $5.5 \; 10^4$  & $1 \; 10^{-5}$  & $3 \; 10^{-9}$ &
$1.7 \; 10^5$ & $2 \; 10^{-2}$ \\
\enddata
\tablenotetext{(a)}{$X(r) = X_0 \exp (-n(r)/n_d)$}
\tablenotetext{(b)}{Model abundance drop: $X(r=20'')/X(r=100'')$}
\end{deluxetable}

\begin{deluxetable}{lcccccc}
\tablecolumns{5}
\tablecaption{Best-Fit Model Parameters for N$_2$H$^+$ and NH$_3$.
\label{tbl-5}}
\tablewidth{0pt}
\tablehead{
\colhead{} & \colhead{N$_2$H$^+$ } &
\colhead{} & \colhead{NH$_3$}  & \colhead{} \\
\colhead{CORE} & \colhead{$X_0$\tablenotemark{(a)}}
& \colhead{$X_0$\tablenotemark{(b)}} & \colhead{$\beta$\tablenotemark{(b)}}
& \colhead{$f$\tablenotemark{(c)}}
}
\startdata
L1498 & $9.0 \; 10^{-11}$ &  $1.0 \; 10^{-8}$  & $1.0$
& $4.1$ \\
L1495 & $2.4 \; 10^{-10}$ & $2.3 \; 10^{-8}$ & $1.0$
& $4.8$ \\
L1400K & $1.3 \; 10^{-10}$ &  $4.0 \; 10^{-9}$ & $0.0$
& $1.0$ \\
L1517B & $9.5 \; 10^{-11}$ & $1.7 \; 10^{-8}$ & $1.0$
& $12$ \\
L1544 & $7.5 \; 10^{-11}$ &  $4.0 \; 10^{-9}$  & $0.3$
& $2.7$ \\
\enddata
\tablenotetext{(a)}{Constant N$_2$H$^+$ abundance.}
\tablenotetext{(b)}{$X($NH$_3) = X_0 (n/n_0)^\beta$}
\tablenotetext{(c)}{Model abundance enhancement: $X(r=20'')/X(r=100'')$.}
\end{deluxetable}

\clearpage

\end{document}